\documentclass[a4paper,11pt]{article}
\usepackage{amsmath,amssymb,graphicx}
\usepackage[hmargin=28mm,vmargin=29mm]{geometry}
\usepackage[numbers,sort&compress]{natbib}
\usepackage[font=small,labelfont=bf]{caption}
\usepackage{tikz,placeins}
\usepackage{microtype}
\usepackage[colorlinks,allcolors=blue,linktocpage]{hyperref}
\numberwithin{equation}{section}

\hypersetup{pdfauthor={Kaifei Ning},
  pdftitle={CP-violating currents from resummed mass mixing in electroweak baryogenesis}}
\usetikzlibrary{arrows.meta,decorations.markings}

\title{CP-violating currents from resummed mass mixing in electroweak baryogenesis}
\author{Kaifei Ning\\[0.5em]
  \small Department of Physics, University of Massachusetts Amherst, Amherst, MA 01003, USA\\
  \small\href{mailto:kning@umass.edu}{kning@umass.edu}}
\date{}
\begin{document}
\maketitle
\begin{abstract}
We calculate the CP-odd current induced by a varying mass phase for two mixing complex scalars and for chiral fermions with a complex Dirac mass. In the closed-time-path formalism, we resum the local mass matrix before expanding in bubble-wall gradients. For hard quasiparticles we use the power counting $L_w^{-1}\ll\gamma_i\ll T$, with $L_w$ the wall thickness and $\gamma_i$ the pole half-width. The resulting current integrals retain the full dependence on the thermal mass shifts and damping widths within our model for the bath self-energies. At vanishing bath chemical potentials, the first-order density is proportional to the temporal phase gradient and the longitudinal flux to the phase gradient normal to the wall. At leading order in the bath couplings, the density can depend on the ratio of the two bath widths, while the longitudinal flux is independent of both. At second order in wall gradients, we independently calculate the mass-mixing and decay sources from the field equations. Their sum reproduces the temporal and spatial parts of the current divergence separately. At equal diagonal scalar masses, resumming the mixing splits the physical poles and removes the pinch that arises in the vacuum expectation value insertion approximation (VIA) as the widths vanish. The leading resummed thermal chiral fermion density remains finite in the formal limit of vanishing common width. At first order in wall gradients, the induced density is concentrated where the phase varies, while the diffusion equation with the VIA source gives a density tail in the symmetric phase.

\end{abstract}

\tableofcontents

\section{Introduction}
\label{sec:introduction}
During a first-order electroweak phase transition, bubbles of the broken phase expand into the surrounding symmetric plasma. An excess of left-handed fermions ahead of a bubble wall can bias weak sphaleron transitions and thereby generate a net baryon number~\cite{Review}. One possible origin of this excess is a complex phase in the particle mass matrix that varies across the wall. Such a phase can induce a chiral current in fermions or a flavor current in mixing scalars. In the scalar case, reactions in the plasma can transfer the induced charge to fermions~\cite{ChungGauge}. The induced charge density depends on how the magnitudes and phases of the mass terms vary across the bubble wall. How accurately existing treatments capture this dependence on the wall profile remains an open question.

Early work described CP-asymmetric reflection from a thin wall and the semiclassical force on particles crossing a thick wall~\cite{Joyce,JoyceThick}. Huet and Nelson included the loss of coherence through scattering and followed the resulting charge through a supersymmetric plasma~\cite{Huet}. Calculations of the current expectation value gave a complementary account of charge production. Comelli, Pietroni and Riotto found a left-handed top current driven by a Higgs phase gradient even at vanishing chemical potentials~\cite{ComelliCurrents1995,ComelliCurrents1996}. Carena et al.\ obtained the first-order stop and Higgsino currents in the minimal supersymmetric Standard Model (MSSM) by resumming the mass insertions induced by the Higgs background to all orders~\cite{CarenaCurrents}. They related the diffusion source to derivatives of the induced density~\cite{CarenaDiffusion}. A sizable density can therefore coexist with a small derivative near its maximum.

In the vacuum expectation value insertion approximation (VIA), the mixing mass enters perturbatively through insertions on unmixed thermal propagators. The relative phase of two insertions gives the conventional CP-violating source at first order in gradients. Early MSSM calculations found enhanced charge production for nearly equal Higgsino and gaugino mass parameters~\cite{Carena}. The closed-time-path (CTP) equations retain the dependence on the wall background between the insertions~\cite{Riotto}. A local expansion of this dependence requires a wall sufficiently thick compared with the damping length~\cite{PostmaRange}. Near degeneracy of the unmixed thermal masses, both the source and the relaxation rate exhibit a resonance~\cite{Lee} that persists at four mass insertions~\cite{PostmaNLO}. The height of the source peak alone therefore gives an incomplete picture of the density available to sphalerons. That density also depends on the relaxation of a pre-existing imbalance~\cite{Weldon,Carosi} and on charge transfer through Yukawa, triscalar and supergauge reactions~\cite{CiriglianoYukawa,ChungGauge,ChungLepton}.

The narrow-width limit exposes a difficulty with the VIA resonance. At equal unmixed energies, a retarded and an advanced pole pinch the real energy contour as their widths tend to zero. Kainulainen identified ambiguities in regulating this pinch with thermal widths for fermions~\cite{KainulainenThermal}. Finite observation times can avoid perturbative pinches in nonequilibrium field theory~\cite{Millington}. For a varying top mass, the conventional VIA and semiclassical sources can yield substantially different baryon asymmetries for the same wall profile~\cite{Cline}. A separate issue concerns the order of the source itself. At leading order in the gradient expansion of the Wigner-space convolutions, Postma, van de Vis and White found that the mass commutator and collision term cancel the first-order VIA source for both scalars and fermions, including after resummation~\cite{PostmaCancel}. What current does the wall induce at first order, and which terms in the field equations account for its divergence at second order?

The Kadanoff--Baym equations illustrate how the gradient order depends on the observable. For a fermion with a complex mass, the CP-odd shift in the dispersion relation contains one gradient and the semiclassical force contains two~\cite{Force,ProkopecI}. This treatment extends to Yukawa collision terms~\cite{ProkopecII} and to flavor-diagonal quasiparticles with thermal self-energies~\cite{Endo}. When several fields mix, the wall can also induce off-diagonal coherence in the mass basis at first order. Konstandin et al.\ computed first-order chargino currents and showed that flavor oscillations shift the peak in charge production~\cite{KonstandinCurrents,KonstandinTransport}. For mixing scalars, Cirigliano et al.\ derived kinetic equations with oscillations and collisions~\cite{Flavor} and found resonant transport when the oscillation length is comparable to the wall thickness~\cite{Oscillations}. At fixed masses, varying the wall thickness changes the relative phase accumulated during a crossing and can move the system through this resonance. Flavor coherence can substantially affect the predicted asymmetry in an extended Higgs sector~\cite{LiFlavor}. To relate this coherence to a flavor current, one must also account for the spacetime dependence of the basis rotation in the current operator.

Here we compute the CP-odd density and longitudinal flux induced by a smooth wall for two mixing complex scalars and for two Weyl fields coupled by a complex Dirac mass. We resum the full local mass matrix in the thermal propagators before expanding in wall gradients. Keeping the wall momentum in a single insertion shows which scales control this expansion [Appendix~\ref{app:gradient_derivation}]. If a same-branch retarded--advanced pair contributes to the observable, $L_w^{-1}\ll\gamma_i$ suffices for the local expansion of its pole factor. For modes with energies of order $T$, the narrow-width assumption $\gamma_i\ll T$ then gives the power counting $L_w^{-1}\ll\gamma_i\ll T$. The local mass resummation permits large mixing and retains the splitting of the physical scalar poles even when the diagonal masses coincide and the widths vanish. The flavor basis identifies the charge coupled to the bath, while the mass basis makes the splitting explicit. Appendix~\ref{app:mass_basis_transport} relates the current in the two bases and shows how first-order coherence can coexist with a vanishing first-order mass-mixing source in a collisionless example.

At vanishing bath chemical potentials, both systems develop a first-order current. Through $\mathcal O(g^0)$ in the bath couplings, the independently calculated mass-mixing and decay sources sum to its divergence at second order in gradients. The temporal and spatial contributions agree separately. This local continuity equation follows from the field equations without a chemical-potential ansatz for the induced charge. Section~\ref{sec:transfer} identifies Yukawa contributions through which the phase can enter the currents of other species.

Section~\ref{sec:numerical} examines how the source and density profiles for stops and chiral top quarks change with masses and widths. For the same moving wall, the leading resummed kernels give the local density and current divergence $\partial_z(j_L^z+v_w n_L)$, while the conventional VIA source drives a diffusion equation that includes relaxation. We test the expansion in the bath coupling by evaluating the resummed wall-center source with the full bath propagators at the same gradient order.

\section{Local observables and power counting}
\label{sec:expansion}
We calculate the left-handed CP-odd density $n_L$ and longitudinal flux $j_L^z$ induced by a planar wall in the plasma rest frame. The bath correlators allow flavor-dependent chemical potentials. We set $\mu_L=\mu_R=0$ to isolate the response to the varying mass phase and hold the temperature and bath self-energy parameters fixed across the wall. The resulting continuity equation relates the divergence of the induced current to two contributions from the field equations, the mass-mixing source $S_L^m$ and the decay source $S_L^d$. The latter arises from the interaction self-energy.

At each observation point $x$ we resum the full local mass matrix in the zeroth-order propagator, allowing arbitrary mixing. Derivatives of that matrix enter as insertions, and a superscript $(n)$ counts the total number of wall derivatives. The induced density and flux begin at first order. Their divergence $\partial_t n_L+\partial_z j_L^z$ therefore contributes at second order, where we compare it with $S_L^m+S_L^d$. The second-order current first contributes to the divergence at third order.

Appendix~\ref{app:gradient_derivation} derives the conditions for this local expansion by retaining the wall momentum in one insertion. At energies of order $T$, the thermal weights vary slowly over the insertion momentum when $L_w^{-1}\ll T$. A retarded and an advanced pole on the same branch are separated by $2\gamma_i$ at zero wall momentum. If the pair contributes to the observable, $L_w^{-1}\ll\gamma_i$ suffices to expand its pole factor. At this stage the four-momentum integrals retain the full widths of the bath model.

To evaluate these integrals analytically, we also expand in the bath coupling $g$. The dispersive self-energies and widths count as $\mathcal O(g^2)$ in thermal units. We take the quasiparticle peaks to be narrow on the thermal scale, $\gamma_i\ll T$. Together with the sufficient locality condition, this gives the regime $L_w^{-1}\ll\gamma_i\ll T$. For distinct branches the analytic expansion is taken at fixed nonzero energy splitting. The sum of their half-widths must be small compared with their local energy separation at the momenta that carry the observable.

We evaluate the current and source kernels through $\mathcal O(g^0)$. Within our self-energy ansatz, we retain the $\mathcal O(g^4)$ shifts that separate poles from numerator zeros because inverse powers of the widths can make the resulting residues contribute to the leading source [Eqs.~\eqref{eq:boson_pole_energy} and~\eqref{eq:boson_pole_width}]. Computing the kernels through $\mathcal O(g^2)$ would also require the residues at the thermal distribution poles. Section~\ref{sec:source_accuracy} tests the expansion in the bath coupling against the full bath propagators at the same gradient order.

\section{Two mixing complex scalars}
\label{sec:scalars}
\subsection{Model and contour conventions}

Consider two complex scalars $\boldsymbol{\Phi}^T=(\Phi_L,\Phi_R)$ with flavor-diagonal bath interactions and a spacetime-dependent mixing mass. Their Lagrangian is
\begin{equation}
  \label{eq:lagrangian_flavor_boson}
\mathcal{L}_{\text{boson}} = \partial_\mu \boldsymbol{\Phi}^\dagger \partial^\mu \boldsymbol{\Phi} - \boldsymbol{\Phi}^\dagger \boldsymbol{M}^2(x) \boldsymbol{\Phi} + (\boldsymbol{J}^\dagger \boldsymbol{\Phi}+\text{c.c.})
\end{equation}
with mass matrix
\begin{equation}
  \boldsymbol{M}^2 = \begin{pmatrix} M_L^2 & v^2(x) e^{-i\alpha(x)} \\v^2(x) e^{i\alpha(x)} & M_R^2 \end{pmatrix},
\end{equation}
Here $M_L^2$ and $M_R^2$ are constant diagonal masses squared. The bath couples through $\boldsymbol J^T=(J_L,J_R)$. All background dependence resides in $v^2(x)e^{-i\alpha(x)}$, as can occur when the scalars couple to the Higgs background across a bubble wall.

On the closed time contour the flavor propagator is
\begin{equation}
\boldsymbol{G}(x,y) = \langle \mathcal{T}^c \boldsymbol{\Phi}(x) \boldsymbol{\Phi}^\dagger(y) \rangle
\end{equation}
where $\mathcal T^c$ orders operators along the contour and the expectation value is taken in the specified initial state. The two contour branches give four components,
\begin{equation}
  \label{eq:defbosonGctp}
\boldsymbol{G}^{t} = \boldsymbol{G}^{++}, ~~\boldsymbol{G}^{\bar{t}} = \boldsymbol{G}^{--}, ~~\boldsymbol{G}^{<} = \boldsymbol{G}^{+-}, ~~\boldsymbol{G}^{>} = \boldsymbol{G}^{-+},
\end{equation}
from which the retarded and advanced propagators follow,
\begin{equation}
  \label{eq:defbosonGra}
  \boldsymbol{G}^{r} = \boldsymbol{G}^{t} - \boldsymbol{G}^{<} = \boldsymbol{G}^{>} - \boldsymbol{G}^{\bar{t}}, ~~~~\boldsymbol{G}^{a} = \boldsymbol{G}^{t} - \boldsymbol{G}^{>} = \boldsymbol{G}^{<} - \boldsymbol{G}^{\bar{t}}.
\end{equation}
We also introduce the spectral density matrix
\begin{equation}
  \label{eq:densityfunction}
  \boldsymbol{\rho} = \boldsymbol{G}^{r} - \boldsymbol{G}^{a} = \boldsymbol{G}^{>} - \boldsymbol{G}^{<}.
\end{equation}
The corresponding one-particle-irreducible self-energy is
\begin{equation}
\boldsymbol{\Pi}(x,y) = -i\langle \mathcal{T}^c \boldsymbol{J}(x) \boldsymbol{J}^\dagger(y) \rangle
\end{equation}
with the same contour assignments as in Eqs.~\eqref{eq:defbosonGctp} and~\eqref{eq:defbosonGra}.

\subsection{Resummed propagator at zeroth order in gradients}
It is convenient to absorb the sign of the backward contour branch into the matrices
\begin{equation}
  \tilde{\boldsymbol{G}} =  \begin{pmatrix} \boldsymbol{G}^t & - \boldsymbol{G}^< \\ \boldsymbol{G}^> & -\boldsymbol{G}^{\bar{t}} \end{pmatrix},
  ~~~~\tilde{\boldsymbol{\Pi}} =  \begin{pmatrix} \boldsymbol{\Pi}^t & - \boldsymbol{\Pi}^< \\ \boldsymbol{\Pi}^> & -\boldsymbol{\Pi}^{\bar{t}} \end{pmatrix}.
\end{equation}
Freezing the background at the observation point $x$ gives the Fourier representation
\begin{equation}
  \label{eq:zero_gradient_momentum_expansion}
    \tilde{\boldsymbol{G}}^{(0)}(x,y) = \int \frac{d^4 k}{(2\pi)^4} e^{-ik(x-y)} \tilde{\boldsymbol{G}}^{(0)}(k,x)
\end{equation}
The second argument of $\tilde{\boldsymbol G}^{(0)}(k,x)$ specifies where the background is evaluated. In a Wigner transform that role is played by the midpoint of the two field arguments.

The local propagator obeys the matrix Dyson equation
\begin{equation}
  \label{eq:inversezerothG}
  \tilde{\boldsymbol{G}}^{(0)}(k,x)^{-1} = \tilde{\boldsymbol{G}}^{(0)}(k)^{-1} + i \tilde{\boldsymbol{M}}^2(x) + i \tilde{\boldsymbol{\Pi}}^{(0)}(k)
\end{equation}
Here $\tilde{\boldsymbol G}^{(0)}(k)$ denotes the free massless propagator and $\tilde{\boldsymbol\Pi}^{(0)}(k)$ the homogeneous bath self-energy. Each matrix acts in the product of contour and flavor space. Locality gives $\tilde{\boldsymbol M}^2=\operatorname{diag}(\boldsymbol M^2,\boldsymbol M^2)$ in contour space.

We impose the Kubo--Martin--Schwinger (KMS) relation separately for each flavor in the free propagator and bath self-energy. The massless propagators are
\begin{equation}
  \begin{aligned}
    \boldsymbol{G}^{(0),r}(k) &= \operatorname{Diag} \left\{\frac{i}{(k^0 + i\epsilon/2)^2 - \boldsymbol{k}^2},\frac{i}{(k^0 + i\epsilon/2)^2 - \boldsymbol{k}^2}\right\},\\
    \boldsymbol{G}^{(0),a}(k) &= \operatorname{Diag} \left\{\frac{i}{(k^0 - i\epsilon/2)^2 - \boldsymbol{k}^2},\frac{i}{(k^0 - i\epsilon/2)^2 - \boldsymbol{k}^2}\right\},\\
    \boldsymbol{G}^{(0),<}(k) &= \boldsymbol{\rho}^{(0)}(k) \boldsymbol{n} \\
    \boldsymbol{G}^{(0),>}(k) &= \boldsymbol{\rho}^{(0)}(k) \left(\boldsymbol{I} + \boldsymbol{n}\right)\\
  \end{aligned}
\end{equation}
where $\boldsymbol{\rho}^{(0)}$ is the free spectral density defined by Eq.~\eqref{eq:densityfunction}, $\boldsymbol I$ is the identity, and
\begin{equation}
  \boldsymbol{n} = \operatorname{Diag} \left\{ n_B(k^0, \mu_L), n_B(k^0, \mu_R)\right\}
\end{equation}  
with $n_B(k^0,\mu_j)=[\exp((k^0-\mu_j)/T)-1]^{-1}$. We parameterize the self-energies as
\begin{equation}
 \begin{aligned}
    \boldsymbol{\Pi}^{(0),r}(k) &= \operatorname{Diag} \left\{\Pi_{h,L} - 2 i (k^0 - \mu_L)\Gamma_L,\Pi_{h,R} - 2 i (k^0 - \mu_R)\Gamma_R\right\},\\
    \boldsymbol{\Pi}^{(0),a}(k) &= \operatorname{Diag} \left\{\Pi_{h,L} + 2 i (k^0 - \mu_L)\Gamma_L,\Pi_{h,R} + 2 i (k^0 - \mu_R)\Gamma_R\right\},\\
    \boldsymbol{\Pi}^{(0),<}(k) &= \left(\boldsymbol{\Pi}^{(0),r} - \boldsymbol{\Pi}^{(0),a}\right)\boldsymbol{n} \\
    \boldsymbol{\Pi}^{(0),>}(k) &= \left(\boldsymbol{\Pi}^{(0),r} - \boldsymbol{\Pi}^{(0),a}\right) \left(\boldsymbol{I} + \boldsymbol{n}\right)\\
  \end{aligned}
\end{equation}
The real constant $\Pi_{h,j}$ shifts the mass squared, and $\Gamma_j$ sets the width in flavor channel $j$. The factor $k^0-\mu_j$ in the absorptive self-energy cancels the real-axis pole of $n_B(k^0,\mu_j)$ in the Wightman components.

The KMS relations give a similarity transformation in contour space for each diagonal flavor channel,
\begin{equation}
  \tilde{\boldsymbol{V}} \tilde{\boldsymbol{G}}^{(0)} \tilde{\boldsymbol{V}}^{-1} = \begin{pmatrix} \boldsymbol{G}^{(0),r} & \boldsymbol{0} \\ \boldsymbol{0} & \boldsymbol{G}^{(0),a}\end{pmatrix},~~~~
  \tilde{\boldsymbol{V}} \tilde{\boldsymbol{\Pi}}^{(0)} \tilde{\boldsymbol{V}}^{-1} = \begin{pmatrix} \boldsymbol{\Pi}^{(0),r} & \boldsymbol{0} \\ \boldsymbol{0} & \boldsymbol{\Pi}^{(0),a}\end{pmatrix}
\end{equation}
with
\begin{equation}
  \tilde{\boldsymbol{V}} = \begin{pmatrix} \boldsymbol{I} + \boldsymbol{n} & - \boldsymbol{n}
  \\ \boldsymbol{I} & - \boldsymbol{I}\end{pmatrix}.
\end{equation}
The mass matrix can be diagonalized by a unitary transformation $\boldsymbol{U}(x)$,
\begin{equation}  
    \boldsymbol{m}^2(x) = \begin{pmatrix} m_1^2 & 0 \\ 0 & m_2^2 \end{pmatrix} = \boldsymbol{U}^\dagger \boldsymbol{M}^2 \boldsymbol{U}, ~~~~\boldsymbol{U}(x) = \begin{pmatrix} \cos\theta(x) & -\sin\theta(x) e^{-i\alpha(x)} \\ \sin\theta(x) e^{i\alpha(x)} & \cos\theta(x) \end{pmatrix},
\end{equation}
where $m_1^2>m_2^2$. We write $c_\theta=\cos\theta$, $s_\theta=\sin\theta$ and $\omega_i=(\boldsymbol k^2+m_i^2)^{1/2}$. With the stated phase convention, $v^2=c_\theta s_\theta(m_1^2-m_2^2)$. The corresponding rotation in contour space is
\begin{equation}  
    \tilde{\boldsymbol{m}}^2(x) = \operatorname{Diag}\left\{{\boldsymbol{m}}^2,{\boldsymbol{m}}^2\right\} = \tilde{\boldsymbol{U}}^\dagger \tilde{\boldsymbol{M}}^2 \tilde{\boldsymbol{U}}, ~~~~\tilde{\boldsymbol{U}}(x) = \operatorname{Diag}\left\{{\boldsymbol{U}},{\boldsymbol{U}}\right\}.
\end{equation}

Rotating to the mass basis and taking the free $i\epsilon$ regulator to zero at finite widths gives
\begin{equation}
  \label{eq:GmEquation}
  \begin{aligned}
    \tilde{\boldsymbol{G}}_m^{(0)}(k,x) \equiv \tilde{\boldsymbol{U}}^\dagger \tilde{\boldsymbol{G}}^{(0)}(k,x)\tilde{\boldsymbol{U}} = \left[-i \left(k^2 - \tilde{\boldsymbol{m}}^2 \right)+ i \tilde{\boldsymbol{U}}^\dagger \tilde{\boldsymbol{V}}^{-1} \operatorname{Diag} \left\{\boldsymbol{\Pi}^{(0),r}, \boldsymbol{\Pi}^{(0),a} \right\} \tilde{\boldsymbol{V}}\tilde{\boldsymbol{U}}\right]^{-1}
  \end{aligned}
\end{equation}
The mass-basis propagator can be written compactly in terms of its retarded,
advanced, and lesser blocks. Suppressing the common arguments $(k,x)$, we define
\begin{equation}
\label{eq:boson_compact_ctp}
\widetilde{\boldsymbol G}^{(0)}_m \simeq
\begin{pmatrix}
 \boldsymbol R+\boldsymbol L & -\boldsymbol L\\
 \boldsymbol R-\boldsymbol A+\boldsymbol L & \boldsymbol A-\boldsymbol L
\end{pmatrix},
\quad
\text{with} \quad
 \boldsymbol R\equiv\boldsymbol G^{(0),r}_m,~
 \boldsymbol A\equiv\boldsymbol G^{(0),a}_m,~
 \boldsymbol L\equiv\boldsymbol G^{(0),<}_m.
\end{equation}
Each block is a $2\times2$ matrix in mass indices. The spectral and statistical
blocks can be written as
\begin{equation}
\label{eq:boson_compact_spectral}
 \boldsymbol R=\frac{i\boldsymbol{\mathcal N}_{+}}{\mathcal D_{+}},
 \qquad
 \boldsymbol A=\frac{i\boldsymbol{\mathcal N}_{-}}{\mathcal D_{-}}, \qquad \boldsymbol L=\boldsymbol R\boldsymbol n-\boldsymbol n\boldsymbol A
              -\boldsymbol R\boldsymbol{\mathcal K}\boldsymbol A,
\end{equation}
Appendix~\ref{app:boson_ctp_coefficients} gives $\boldsymbol{\mathcal N}_{\pm}$, $\mathcal D_{\pm}$ and $\boldsymbol{\mathcal K}$ explicitly and shows how the numerator zeros and propagator poles depend on the thermal widths and chemical potentials.
We expand the pole positions and numerator zeros through $\mathcal O(g^4)$ and keep their factors unexpanded in the propagator. For a common chemical
potential, $\boldsymbol{\mathcal K}=0$ and
$\boldsymbol L=n_B(\boldsymbol R-\boldsymbol A)$. For unequal chemical potentials we use the full expression for the lesser propagator $\boldsymbol L$.
Rotation by $\tilde{\boldsymbol U}$ gives the flavor-basis components entering the current.

\subsection{Charge density and longitudinal current}
For the left scalar the charge current is
\begin{equation}
  \label{eq:defbosoncurrent}
  \begin{aligned}
    j_L^\mu(x) &\equiv i \left\langle :  \Phi_L^\ast (\partial^\mu \Phi_L) - (\partial^\mu \Phi_L^\ast)\Phi_L :\right\rangle \\
    &= \frac{i}{2} \left\langle  \Phi_L^\ast (\partial^\mu \Phi_L) - (\partial^\mu \Phi_L^\ast)\Phi_L\right\rangle - \frac{i}{2} \left\langle  \Phi_L (\partial^\mu \Phi_L^\ast) - (\partial^\mu \Phi_L)\Phi_L^\ast\right\rangle \\
    &= - \operatorname{Im} \lim_{y\to x}\partial_{x_\mu}\left( G_{LL}^<(x,y)+G_{LL}^>(x,y)\right)
  \end{aligned}
\end{equation}
We evaluate its CP-odd part at $\mu_L=\mu_R=0$ using the coincident two-point function. Appendix~\ref{app:gradient_derivation} derives the gradient insertions used below.

\subsubsection{Homogeneous background}
For a homogeneous background, a constant rephasing of the scalar fields removes $\alpha$ from the mass matrix and leaves the flavor-diagonal bath self-energies and the current operator invariant. At $\mu_L=\mu_R=0$, the CP-odd current therefore vanishes to all orders in the bath couplings,
\begin{equation}
  \label{eq:zerothcpvcurrent}
  {}^{\text{CPV}(0)}j_L^{\mu}(x)
  = \operatorname{Re}\int\frac{d^4 k}{(2\pi)^4}\, k^\mu
    \left(G_{LL}^{(0),<}(k,x)+G_{LL}^{(0),>}(k,x)\right) = 0 .
\end{equation}
The Wightman functions fall as $\mathcal O((k^0)^{-3})$ at large energy away from their thermal poles, allowing the contour to be closed. The quasiparticle residues reproduce this zero through $\mathcal O(g^0)$, the order retained in the pole calculation.

\subsubsection{First-order response}
For the planar background the first-order charge density is
\begin{equation}
  \label{eq:fonL}
  \begin{aligned}
    {}^{\text{CPV}(1)}n_L(x)
    &= v^2 \partial_t\alpha\,\operatorname{Im}\sum_{\lambda = <,>}\int \frac{d^4k}{(2\pi)^4}
      \Bigg[\partial_{k^0}\!\left(k^0 G^r_{LL}\right)e^{-i\alpha}G^\lambda_{RL}
      - \partial_{k^0}\!\left(k^0 G^r_{LR}\right)e^{i\alpha}G^\lambda_{LL} \\
    &\qquad\qquad
      - k^0 G^\lambda_{LL}e^{-i\alpha}\partial_{k^0}G^a_{RL}
      + k^0 G^\lambda_{LR}e^{i\alpha}\partial_{k^0}G^a_{LL}\Bigg] \\
    &\simeq -v^2 \partial_t\alpha \int \frac{d^3k}{(2\pi)^3}
      \frac{c_\theta s_\theta}{(\omega_1^2-\omega_2^2)^2}
      \Bigg[
      \frac{3\omega_1^2+\omega_2^2}{\omega_1}\coth\!\left(\frac{\omega_1}{2T}\right)
      - \frac{\omega_1^2+3\omega_2^2}{\omega_2}\coth\!\left(\frac{\omega_2}{2T}\right) \\
    &\qquad
      + \frac{(\omega_1^2-\omega_2^2)(\Gamma_L-\Gamma_R)}{2T}
      \left(
        \frac{c_\theta^2\operatorname{csch}^2\!\left(\frac{\omega_1}{2T}\right)}
             {\Gamma_L c_\theta^2+\Gamma_R s_\theta^2}
        + \frac{s_\theta^2\operatorname{csch}^2\!\left(\frac{\omega_2}{2T}\right)}
             {\Gamma_L s_\theta^2+\Gamma_R c_\theta^2}
      \right)
      \Bigg].
  \end{aligned}
\end{equation}
Only the temporal phase gradient enters the density at this order. In this equation and Eq.~\eqref{eq:fojLz}, the four-momentum integral retains the widths, and $\simeq$ denotes its leading term at fixed mass splitting.

Throughout, we drop total $k_3$ derivatives under the longitudinal
momentum integral. We verified this step explicitly for the first-order
CP-violating spatial current, which is given by
\begin{equation}
  \label{eq:fojLz}
  \begin{aligned}
    {}^{\text{CPV}(1)}j_L^z(x)
    &= -v^2 \partial_z\alpha\,\operatorname{Im}\sum_{\lambda = <,>}\int \frac{d^4k}{(2\pi)^4}
      \Bigg[\partial_{k^3}\!\left(k^3 G^r_{LL}\right)e^{-i\alpha}G^\lambda_{RL}
      - \partial_{k^3}\!\left(k^3 G^r_{LR}\right)e^{i\alpha}G^\lambda_{LL} \\
    &\qquad\qquad
      - k^3 G^\lambda_{LL}e^{-i\alpha}\partial_{k^3}G^a_{RL}
      + k^3 G^\lambda_{LR}e^{i\alpha}\partial_{k^3}G^a_{LL}\Bigg] \\
    &\simeq -v^2 \partial_z\alpha \int \frac{d^3k}{(2\pi)^3}
      \frac{c_\theta s_\theta}{(\omega_1^2-\omega_2^2)^2} \\
    &\quad \times \Bigg[
      (\omega_1^2-\omega_2^2)\!
      \left(\frac{\coth(\omega_1/2T)}{\omega_1}
           +\frac{\coth(\omega_2/2T)}{\omega_2}\right)
      -8T\log\!\left(\frac{\sinh(\omega_1/2T)}{\sinh(\omega_2/2T)}\right)
      \Bigg].
  \end{aligned}
\end{equation} 
The spatial current is proportional to $\partial_z\alpha$. Both the density and the longitudinal flux vanish as $s_\theta\to0$ and can remain finite when $\Gamma_L=\Gamma_R\to0$. The divergence of the first-order current
\begin{equation}
  S^{\text{CPV}}(x) \equiv \lim_{\mu_L, \mu_R \to0}\partial_\mu j^\mu(x)
\end{equation} 
therefore begins at second order in gradients. We next evaluate the two contributions to this divergence independently.

\subsection{Sources from the scalar field equations}
\label{sec:bosonsource2nd}
Differentiating the current in Eq.~\eqref{eq:defbosoncurrent} gives
\begin{equation}
  \label{eq:EOMsource}
  \begin{aligned}
    S_L(x) &= \frac{i}{2} \left\langle  \Phi_L^\ast (\square \Phi_L) - (\square\Phi_L^\ast)\Phi_L\right\rangle - \frac{i}{2} \left\langle  \Phi_L (\square \Phi_L^\ast) - (\square \Phi_L)\Phi_L^\ast\right\rangle \\
    &= \frac{i}{2}\left\langle (\square \Phi_L)\Phi_L^\ast\right\rangle  +\frac{i}{2} \left\langle  \Phi_L^\ast (\square \Phi_L)\right\rangle + \text{c.c.}
  \end{aligned}
\end{equation}
Variation with respect to $\boldsymbol{\Phi}^\dagger$ gives
\begin{equation}
  \square \boldsymbol\Phi(x) = -\boldsymbol{M}^2(x) \boldsymbol{\Phi} + \boldsymbol{J}
\end{equation}
and therefore,
\begin{equation}
  \begin{aligned}
  S_L(x) &=-\frac{i}{2}v^2 e^{-i\alpha }\left\langle \Phi_R \Phi_L^\ast\right\rangle  -\frac{i}{2} v^2 e^{-i\alpha } \left\langle  \Phi_L^\ast \Phi_R\right\rangle  +  \frac{i}{2}\left\langle J_L \Phi_L^\ast\right\rangle+  \frac{i}{2}\left\langle \Phi_L^\ast J_L \right\rangle  + \text{c.c.} \\
  &= \operatorname{Im}\lim_{y\to x} \left[v^2 e^{-i\alpha } \left(G_{RL}^>(x,y) + G_{RL}^<(x,y) \right) - \left(D_{LL}^>(x,y) + D_{LL}^<(x,y)\right)\right]
  \end{aligned}
\end{equation}
Here $\boldsymbol D(x,y)=\langle\mathcal T^c\boldsymbol J(x)\boldsymbol\Phi^\dagger(y)\rangle$. The terms involving $G^\lambda_{RL}$ define the mass-mixing source $S_L^m$, and those involving $D^\lambda_{LL}$ define the decay source $S_L^d$.
\subsubsection{Mass-mixing source}
\label{sec:massmixingsource}
We take $\mu_L,\mu_R\to0$ throughout this section. The zeroth- and
first-order mass-mixing sources vanish at the order retained here.
Writing $\partial_\sigma=\partial/\partial x^\sigma$, their integrals are
\begin{equation}
    {}^{\text{CPV}(0)}S_L^m(x) = v^2 \operatorname{Im} \int \frac{d^4 k}{(2\pi)^4} e^{-i\alpha }\left( G_{RL}^{(0),<}(k,x)+G_{RL}^{(0),>}(k,x)\right)
\end{equation}
and
\begin{equation}
  \begin{aligned}
    {}^{\text{CPV}(1)}S_L^m(x)
    &= -v^4\sum_{\sigma=0,3}(-1)^\sigma(\partial_\sigma\alpha)
       \operatorname{Re}\sum_{\lambda=<,>}\int\frac{d^4k}{(2\pi)^4}\\
    &\quad\times\Big[e^{-2i\alpha}\big((\partial_{k^\sigma}G^r_{RL})G^\lambda_{RL}
                         -G^\lambda_{RL}\partial_{k^\sigma}G^a_{RL}\big)\\
    &\qquad
       -\big((\partial_{k^\sigma}G^r_{RR})G^\lambda_{LL}
                   -G^\lambda_{RR}\partial_{k^\sigma}G^a_{LL}\big)
       \Big].
  \end{aligned}
\end{equation}
The factor $(-1)^\sigma$ converts $\partial_{k_\sigma}$ to
$\partial_{k^\sigma}$ for $\sigma=0,3$. The spatial first-order
integrand is odd in $k^3$ and integrates to zero. At leading order in the bath coupling, the mass-mixing
source starts at second order in gradients,
\begin{equation}
  \label{eq:masssource2nd}
  \begin{aligned}
    &{}^{\text{CPV}(2)}S_L^{m}(x) \\
    &= \sum_{\sigma=0,3}\frac{1}{2}v^2  \left(v^2\partial_{x^\sigma}^2\alpha + 2 \partial_{x^\sigma} v^2 \partial_{x^\sigma}\alpha \right) \operatorname{Im} \sum_{\lambda = <,>}\int\frac{d^4 k}{(2\pi)^4} e^{-i\alpha}\bigg[\partial_{k^\sigma}^2  G^r_{RL} e^{-i\alpha} G^\lambda_{RL} \\
    &\qquad\quad {}- \partial_{k^\sigma}^2 G^r_{RR} e^{i\alpha} G^\lambda_{LL} + G^\lambda_{RL} e^{-i\alpha} \partial_{k^\sigma}^2 G^a_{RL} -  G^\lambda_{RR} e^{i\alpha} \partial_{k^\sigma}^2 G^a_{LL}\bigg] \\
    & +\sum_{\sigma=0,3} 2v^4 \partial_{x^\sigma} v^2 \partial_{x^\sigma}\alpha \operatorname{Re} \sum_{\lambda = <,>}\int\frac{d^4 k}{(2\pi)^4} e^{-i\alpha}\bigg[ \\
    &\qquad e^{-2i\alpha}\big(\partial_{k^\sigma} G^r_{RL}  G^\lambda_{RL} \partial_{k^\sigma} G^a_{RL}+ \partial_{k^\sigma} G^r_{RL} G^r_{RL} \partial_{k^\sigma} G^\lambda_{RL} + \partial_{k^\sigma} G^\lambda_{RL}  G^a_{RL} \partial_{k^\sigma} G^a_{RL}\big) \\
    &-e^{2i\alpha}\left(\partial_{k^\sigma} G^r_{RR}  G^\lambda_{LR} \partial_{k^\sigma} G^a_{LL}+\partial_{k^\sigma} G^r_{RR}  G^r_{LR}\partial_{k^\sigma} G^\lambda_{LL}+\partial_{k^\sigma} G^\lambda_{RR}  G^a_{LR} \partial_{k^\sigma} G^a_{LL}\right)\bigg]
  \end{aligned}
\end{equation}

\subsubsection{Decay source}

The mixed correlator obeys the position-space convolution
\begin{equation}
  \lim_{y\to x}\tilde{\boldsymbol{D}}(x,y) = -\lim_{y\to x}\int d^4 \omega \tilde{\boldsymbol{\Pi}}^{(0)}(x,\omega) \tilde{\boldsymbol{G}}(\omega,y)
\end{equation}
and
\begin{equation}
  \label{eq:Dsourceposition}
  \begin{aligned}
    {}^{\text{CPV}}S_L^{d}(x) &= \lim_{y\to x}\operatorname{Im}\sum_{\lambda = <,>}\int d^4 \omega \Pi_{LL}^{(0),r}(x,\omega) G_{LL}^\lambda(\omega,y) + \Pi_{LL}^{(0),\lambda}(x,\omega) G_{LL}^a(\omega,y).
  \end{aligned}
\end{equation}
The retarded self-energy gives the first contribution,
\begin{equation}
  \begin{aligned}
      \left.{}^{\mathrm{CPV}}S_L^d(x)\right|_1 &=\lim_{y\to x}\operatorname{Im}\int d^4 \omega \left(\int \frac{d^4 k }{(2\pi)^4} (\Pi_{h,L} - 2i k^0 \Gamma_L) e^{-ik(x-\omega)}\right) \sum_{\lambda = <,>}G_{LL}^\lambda(\omega,y)\\
      &= \lim_{y\to x}\operatorname{Im}\int d^4  \omega (\Pi_{h,L}  - 2\Gamma_L \partial_{\omega^0})\delta^4(x-\omega)\sum_{\lambda = <,>}G_{LL}^\lambda(\omega,y)\\
      &= 2\Gamma_L \int d^4  \omega   \delta^4(x-\omega)\lim_{y\to x}\operatorname{Im} \sum_{\lambda = <,>}\partial_{\omega^0} G_{LL}^\lambda(\omega,y)\\
      &= -2\Gamma_L n_L^{\text{CPV}}(x),
  \end{aligned}
\end{equation}
where we used integration by parts and the reality of $\Pi_{h,L}$ and the coincident diagonal Wightman functions.
The second term defines $n_L^{\mathrm{eq,CPV}}(x)$ through the convolution
\begin{equation}
  \label{eq:boson_decay_gain}
  \begin{aligned}
    \left.{}^{\mathrm{CPV}}S_L^d(x)\right|_2
      &=-4\Gamma_L\operatorname{Re}
        \int d^4\omega\int\frac{d^4k}{(2\pi)^4}
        k^0(1+2n_B(k^0))e^{-ik(x-\omega)}
        G_{LL}^a(\omega,x)\\
      &\equiv 2\Gamma_L n_L^{\mathrm{eq,CPV}}(x),
  \end{aligned}
\end{equation}
where the CP-odd part of the convolution is understood.
In a homogeneous background, $n_L^{\mathrm{eq,CPV}}$ reduces to the
spectral integral weighted by $k^0(1+2n_B)$.
Through $\mathcal O(g^0)$, the higher-gradient contributions to the
gain term vanish upon contour integration.
At vanishing chemical potentials, the decay source satisfies
\begin{equation}
  {}^{\mathrm{CPV}}S_L^d(x)
  =-2\Gamma_L\left(n_L^{\mathrm{CPV}}(x)-n_L^{\mathrm{eq,CPV}}(x)\right)
\end{equation}
with the convolution definition of $n_L^{\mathrm{eq,CPV}}$.
A similar form may hold at finite chemical potentials, where the factor $k^0-\mu_L$ in the self-energy must be taken into account when evaluating the gain term.

For the gradient expansion of the decay source, the zeroth-order mixed correlator is
\begin{equation}
  \tilde{\boldsymbol{D}}^{(0)}(k,x) = -\tilde{\boldsymbol{\Pi}}^{(0)}(k,x) \tilde{\boldsymbol{G}}^{(0)}(k,x),
\end{equation}
The same insertions as in Section~\ref{sec:massmixingsource} give the zeroth- and first-order sources, both of which vanish at the retained order,
\begin{equation}
  {}^{\text{CPV}(0)}S_L^d(x) = -\operatorname{Im} \int \frac{d^4 k}{(2\pi)^4} \left( D_{LL}^{(0),<}(k,x)+D_{LL}^{(0),>}(k,x)\right),
\end{equation}
and
\begin{equation}
  \begin{aligned}
    {}^{\text{CPV}(1)}S_L^d(x)
    &= v^2\sum_{\sigma=0,3}(-1)^\sigma(\partial_\sigma\alpha)
       \operatorname{Re}\sum_{\lambda=<,>}\int\frac{d^4k}{(2\pi)^4}
       \Big[
       e^{-i\alpha}\big((\partial_{k^\sigma}D^r_{LL})G^\lambda_{RL}
                       -D^\lambda_{LL}\partial_{k^\sigma}G^a_{RL}\big)\\
    &\qquad
       -e^{i\alpha}\big((\partial_{k^\sigma}D^r_{LR})G^\lambda_{LL}
                       -D^\lambda_{LR}\partial_{k^\sigma}G^a_{LL}\big)
       \Big].
  \end{aligned}
\end{equation}
Using the exact, unexpanded scalar propagator of Eq.~\eqref{eq:GmEquation}, we verified in \textbf{Mathematica} that the first-order mass-mixing and decay sources cancel to all orders in $g$ for the bath self-energies used here. At second order in gradients, the decay source through $\mathcal O(g^0)$ is
\begin{equation}
  \label{eq:decaysource2nd}
  \begin{aligned}
    &{}^{\text{CPV}(2)}S_L^{d}(x) \\
    &= -\sum_{\sigma=0,3}\frac{1}{2} \left(v^2\partial_{x^\sigma}^2\alpha + 2 \partial_{x^\sigma} v^2 \partial_{x^\sigma}\alpha \right) \operatorname{Im} \sum_{\lambda = <,>}\int\frac{d^4 k}{(2\pi)^4} \bigg[\partial_{k^\sigma}^2  D^r_{LL} e^{-i\alpha} G^\lambda_{RL} \\
    &\qquad\quad {}- \partial_{k^\sigma}^2 D^r_{LR} e^{i\alpha} G^\lambda_{LL} + D^\lambda_{LL} e^{-i\alpha} \partial_{k^\sigma}^2 G^a_{RL} -  D^\lambda_{LR} e^{i\alpha} \partial_{k^\sigma}^2 G^a_{LL}\bigg] \\
    & -\sum_{\sigma=0,3} 2v^2 \partial_{x^\sigma} v^2 \partial_{x^\sigma}\alpha \operatorname{Re} \sum_{\lambda = <,>}\int\frac{d^4 k}{(2\pi)^4} \bigg[ \\
    &\qquad e^{-2i\alpha}\big(\partial_{k^\sigma} D^r_{LL}  G^\lambda_{RL} \partial_{k^\sigma} G^a_{RL}+ \partial_{k^\sigma} D^r_{LL} G^r_{RL} \partial_{k^\sigma} G^\lambda_{RL} + \partial_{k^\sigma} D^\lambda_{LL}  G^a_{RL} \partial_{k^\sigma} G^a_{RL}\big) \\
    &-e^{2i\alpha}\left(\partial_{k^\sigma} D^r_{LR}  G^\lambda_{LR} \partial_{k^\sigma} G^a_{LL}+\partial_{k^\sigma} D^r_{LR}  G^r_{LR}\partial_{k^\sigma} G^\lambda_{LL}+\partial_{k^\sigma} D^\lambda_{LR}  G^a_{LR} \partial_{k^\sigma} G^a_{LL}\right)\bigg].
  \end{aligned}
\end{equation}
The explicit results of Eqs.~\eqref{eq:masssource2nd} and \eqref{eq:decaysource2nd} are given in Appendix~\ref{app:boson_second_order_sources}.

\subsection{Continuity equation and scalar mass splitting}
Equations~\eqref{eq:fonL} and~\eqref{eq:fojLz} determine the two terms on the left-hand side of the CP-odd continuity equation,
\begin{equation}
  \partial_t\!\left({}^{\mathrm{CPV}(1)}n_L\right)
 +\partial_z\!\left({}^{\mathrm{CPV}(1)}j_L^z\right)
 ={}^{\mathrm{CPV}(2)}S_L^m+{}^{\mathrm{CPV}(2)}S_L^d .
\end{equation}
Differentiation acts on $v^2$ and $\alpha$ as well as on the local eigenfrequencies and mixing angle. The temporal and spatial contributions reproduce separately the kernels obtained from Eqs.~\eqref{eq:masssource2nd} and~\eqref{eq:decaysource2nd}. The scalar mass matrix fixes the splitting between the physical masses even in the absence of damping,
\begin{equation}
 \omega_1^2-\omega_2^2=m_1^2-m_2^2
   =\sqrt{(M_L^2-M_R^2)^2+4v^4}.
\end{equation}
Even at $M_L^2=M_R^2$, nonzero mixing gives a splitting $2v^2$. The eigenstates are degenerate only if the diagonal masses are equal and $v^2=0$, in which case the phase drops out of the mass matrix. This splitting keeps the resummed source free of the VIA pinch singularity at degenerate diagonal masses, even as the widths vanish. Width ratios can still affect the magnitude of the density. Appendix~\ref{app:mass_basis_transport} relates this local mass source to off-diagonal coherence in the mass basis, including at small splitting.

For a wall at rest in the plasma, $\partial_t\alpha=0$ and the first-order density vanishes. A spatial phase gradient can still induce a flavor current whose divergence balances local flavor conversion, $\partial_z j_L^{z(1)}=S_L^{m(2)}+S_L^{d(2)}$. If the current vanishes on both sides of the wall, the integrated source is zero.

\section{A Dirac fermion with a complex mass}
\label{sec:dirac}
\subsection{Model and chiral propagators}
We consider two Weyl fields coupled by a Dirac mass $m_D(x)e^{-i\alpha(x)}$. Their bath interactions are diagonal in chirality. The Lagrangian is
\begin{equation}
\mathcal{L} = i\bar{\boldsymbol{\psi}}\not\!\partial \boldsymbol{\psi} - \left( \psi_L^\dagger m_D e^{-i\alpha} \psi_R - j_L^\dagger \psi_L - j_R^\dagger \psi_R + \text{h.c.}\right)
\end{equation}
where $\psi_L$ and $\psi_R$ are the chiral components of $\boldsymbol\psi$. The operators $j_L$ and $j_R$ couple the corresponding fields to the bath. Defining
\begin{equation}
  \boldsymbol{j} = \begin{pmatrix} j_R \\ j_L \end{pmatrix}, \qquad  \qquad
  \boldsymbol{M} = \begin{pmatrix} m_D e^{i\alpha} & 0 \\ 0 & m_D e^{-i\alpha}\end{pmatrix}
\end{equation}
we can write the Lagrangian in the form
\begin{equation}
  \mathcal{L}_{\text{fermion}} = i\bar{\boldsymbol{\psi}}\not\!\partial \boldsymbol{\psi} - \bar{\boldsymbol{\psi}} \boldsymbol{M} \boldsymbol{\psi} + \left(\bar{\boldsymbol{j}}\boldsymbol{\psi} + \bar{\boldsymbol{\psi}} \boldsymbol{j} \right).
\end{equation}
The contour propagator and self-energy are
\begin{equation}
  \boldsymbol{S}(x,y) = \langle \mathcal{T}^c \boldsymbol{\psi}(x) \bar{\boldsymbol{\psi}}(y)\rangle, \qquad \boldsymbol{\Pi}(x,y) = -i\langle \mathcal{T}^c \boldsymbol{j}(x) \bar{\boldsymbol{j}}(y) \rangle
\end{equation}
with the branch assignments of Eq.~\eqref{eq:defbosonGctp}. Contour ordering includes the minus sign for exchanging fermion fields. With this convention the retarded, advanced and spectral combinations obey Eqs.~\eqref{eq:defbosonGra} and~\eqref{eq:densityfunction}.

\subsection{Resummed propagator at fixed helicity}
We again use the CTP correlation matrices
\begin{equation}
  \tilde{\boldsymbol{S}} =  \begin{pmatrix} \boldsymbol{S}^t & - \boldsymbol{S}^< \\ \boldsymbol{S}^> & -\boldsymbol{S}^{\bar{t}} \end{pmatrix},
  \qquad \tilde{\boldsymbol{\Pi}} =  \begin{pmatrix} \boldsymbol{\Pi}^t & - \boldsymbol{\Pi}^< \\ \boldsymbol{\Pi}^> & -\boldsymbol{\Pi}^{\bar{t}} \end{pmatrix}, \qquad 
  \tilde{\boldsymbol{M}} =  \begin{pmatrix} \boldsymbol{M} & \boldsymbol{0} \\ \boldsymbol{0} & \boldsymbol{M} \end{pmatrix}.
\end{equation}
The momentum-space propagator at zeroth order in gradients satisfies
\begin{equation}
  \label{eq:resumfermionpropagator}
  \tilde{\boldsymbol{S}}^{(0)}(k,x)^{-1} = \tilde{\boldsymbol{S}}^{(0)}(k)^{-1} + i \tilde{\boldsymbol{M}}(x) + i \tilde{\boldsymbol{\Pi}}^{(0)}(k).
\end{equation}
We assume the free propagator and self-energy satisfy the KMS relations in
each chiral channel $I=L,R$, with occupation
$n_{F,I}=n_F(k^0,\mu_I)=[\exp((k^0-\mu_I)/T)+1]^{-1}$.
In each channel, the free propagator and the self-energy are reconstructed
from their diagonal retarded and advanced blocks,
\begin{equation}
  \begin{aligned}
    \tilde{\boldsymbol S}_{II}^{(0)}(k)
      &= V_F(n_{F,I})^{-1}
         \begin{pmatrix}\boldsymbol S_{II}^{(0),r}(k) & \boldsymbol 0\\
                         \boldsymbol 0 & \boldsymbol S_{II}^{(0),a}(k)\end{pmatrix}
         V_F(n_{F,I}),
      \qquad V_F(n)=\begin{pmatrix}1-n&n\\1&-1\end{pmatrix},\\
    \tilde{\boldsymbol\Pi}_{II}^{(0)}(k)
      &= V_F(n_{F,I})^{-1}
         \begin{pmatrix}\boldsymbol\Pi_{II}^{(0),r}(k) & \boldsymbol 0\\
                         \boldsymbol 0 & \boldsymbol\Pi_{II}^{(0),a}(k)\end{pmatrix}
         V_F(n_{F,I}).
  \end{aligned}
\end{equation}
These channel blocks supply the free propagator and self-energy in
Eq.~\eqref{eq:resumfermionpropagator}. We drop the free $i\epsilon$
prescriptions in the resummed propagator in the presence of finite decay
widths.

Following Ref.~\cite{PostmaCancel}, we introduce the helicity projectors
\begin{equation}
  P^\pm = \frac{1}{2} \left({1} \pm \vec{\sigma} \cdot \hat{k}\right)
\end{equation} 
where $\hat{k}=\boldsymbol k/k$, $k=|\boldsymbol k|$, and $\vec\sigma$ are the Pauli matrices. The projectors obey $P^+P^-=0$, $(P^\pm)^2=P^\pm$ and $P^++P^-=\boldsymbol1$. With $K_\pm=k^0\pm k$,
\begin{equation}
  k_\mu \boldsymbol{\gamma^\mu} = \begin{pmatrix} {0} & K_- P^++K_+ P^- \\  K_+ P^++K_- P^- & {0}\end{pmatrix}.
\end{equation}
For the dispersive and absorptive components we take
\begin{equation}
  \begin{aligned}
    {\Pi}_{LL}^h &= \Pi_{LL,+}^h P^+ +\Pi_{LL,-}^hP^-, \qquad {\Pi}_{RR}^h = \Pi_{RR,-}^h P^+ +\Pi_{RR,+}^hP^- \\
    {\Pi}_{LL}^\mathcal{A} &= 2\Gamma_L (P^++P^-), \qquad \qquad {\Pi}_{RR}^\mathcal{A} = 2\Gamma_R (P^++P^-)
  \end{aligned}
\end{equation}
where $\Pi^h_{II,\pm}=a_I K_\pm+b_I$ consists of a term linear in $K_\pm$ and a constant shift $b_I$ for each chirality $I=L,R$. In the plasma rest frame $a_I$ multiplies $\not\!k$ and $b_I$ multiplies $\not\!u$, with $u^\mu$ the plasma four-velocity.
We take $\Gamma_I$, $a_I$ and $b_I$ to be momentum independent and to
include their $\mathcal{O}(g^2)$ scaling.
The retarded and advanced self-energy matrices are given by $\boldsymbol{\Pi}^{r,a} = \boldsymbol{\Pi}^{h}\mp i \boldsymbol{\Pi}^{\mathcal{A}}$.

\begin{samepage}
Suppressing $(k,x)$, the resummed CTP propagator of
Eq.~\eqref{eq:resumfermionpropagator} takes the compact form
\begin{equation}
  \label{eq:fermion_compact_ctp}
  \tilde{\boldsymbol{S}}^{(0)} \simeq
  \sum_{h=\pm1}
  \begin{pmatrix}
    \boldsymbol{R}_h+\boldsymbol{L}_h & -\boldsymbol{L}_h\\
    \boldsymbol{R}_h-\boldsymbol{A}_h+\boldsymbol{L}_h &
    \boldsymbol{A}_h-\boldsymbol{L}_h
  \end{pmatrix}
  \otimes P^{h}.
\end{equation}
The outer matrix is ordered by CTP branch. Each block has rows
$(\psi_L,\psi_R)$ and columns $(\psi_R^\dagger,\psi_L^\dagger)$.
The retarded, advanced, and lesser blocks are
\begin{equation}
  \label{eq:fermion_compact_blocks}
  \begin{aligned}
    \boldsymbol{R}_h &= \frac{i\boldsymbol{\mathcal{N}}_{h,+}}{\mathcal D_{h,+}},
    \qquad
    \boldsymbol{A}_h = \frac{i\boldsymbol{\mathcal{N}}_{h,-}}{\mathcal D_{h,-}},\\
    \boldsymbol{L}_h &= -\frac{n_{F,L}+n_{F,R}}{2}
                         (\boldsymbol{R}_h-\boldsymbol{A}_h)
                       -\boldsymbol{R}_h\boldsymbol{\mathcal K}_F\boldsymbol{A}_h.
  \end{aligned}
\end{equation}
\end{samepage}
Appendix~\ref{app:fermion_ctp_coefficients} gives the explicit numerators,
factored denominators, and statistical kernel.

\subsection{Left-handed charge density and longitudinal current}
We define the left-handed current by the antisymmetrized fermion bilinear,
\begin{equation}
  \begin{aligned}
    j_L^\mu &\equiv\frac{1}{2} \operatorname{Tr} \left[\left\langle {\psi}_L^\dagger \bar{\sigma}^\mu \psi_L \right\rangle - \left\langle {\psi}_L( \bar{\sigma}^\mu)^T \psi_L^\dagger \right\rangle\right] \\
    &= -\frac{1}{2} \operatorname{Tr} \lim_{x\to y}\left[ S^<_{LL}(x,y) \bar{\sigma}^\mu + S^>_{LL}(x,y) \bar{\sigma}^\mu\right]
  \end{aligned}
\end{equation}
Here $\bar\sigma^\mu=(\boldsymbol1,-\vec\sigma)$. We insert the propagator of Eqs.~\eqref{eq:fermion_compact_ctp} and~\eqref{eq:fermion_compact_blocks} into the gradient expansion of Appendix~\ref{app:gradient_derivation}.

At $\mu_L=\mu_R=0$ the homogeneous CP-odd current vanishes. At first order in gradients the left-handed projection of Eq.~\eqref{eq:first_order_gradient} gives the four-momentum integral below. The final radial kernel is its leading bath-coupling term.
\begin{equation}
  \label{eq:fermionfonL}
  \begin{aligned}
    {}^{\text{CPV}}n_L(x)
    &= \frac{m_D}{2}\partial_t\alpha\,\operatorname{Im}
       \sum_{\lambda=<,>}\int\frac{d^4k}{(2\pi)^4}\operatorname{Tr}\Big[
       e^{i\alpha}\big((\partial_{k^0}S^r_{LR})S^\lambda_{LL}
                       -S^\lambda_{LR}\partial_{k^0}S^a_{LL}\big)\\
    &\qquad
       -e^{-i\alpha}\big((\partial_{k^0}S^r_{LL})S^\lambda_{RL}
                       -S^\lambda_{LL}\partial_{k^0}S^a_{RL}\big)\Big]\\
    &\simeq m_D^2\partial_t\alpha\int\frac{d^3k}{(2\pi)^3}
       \Bigg[\frac{\tanh(\omega/2T)}{2\omega^3}\\
    &\qquad
       -\frac{\operatorname{sech}^2(\omega/2T)}{4T\omega^2}
        \frac{(\Gamma_L-\Gamma_R)
              [2\Gamma_R k^2+(\Gamma_L+\Gamma_R)m_D^2]}
             {4\Gamma_L\Gamma_R k^2+(\Gamma_L+\Gamma_R)^2m_D^2}
       \Bigg].
  \end{aligned}
\end{equation}
The spatial current at the same order is
\begin{equation}
  \label{eq:fermionfojLz}
  \begin{aligned}
    {}^{\text{CPV}}j_L^z(x)
    &= \frac{m_D}{2}\partial_z\alpha\,\operatorname{Im}
       \sum_{\lambda=<,>}\int\frac{d^4k}{(2\pi)^4}\operatorname{Tr}\Big\{
       \big[e^{i\alpha}\big((\partial_{k^3}S^r_{LR})S^\lambda_{LL}
                          -S^\lambda_{LR}\partial_{k^3}S^a_{LL}\big)\\
    &\qquad
       -e^{-i\alpha}\big((\partial_{k^3}S^r_{LL})S^\lambda_{RL}
                       -S^\lambda_{LL}\partial_{k^3}S^a_{RL}\big)\big]\sigma^3
       \Big\}\\
    &\simeq m_D^2\partial_z\alpha\int\frac{d^3k}{(2\pi)^3}
       \Bigg[\frac{\operatorname{sech}^2(\omega/2T)}{4T\omega^2}
             -\frac{\tanh(\omega/2T)}{2\omega^3}
       \Bigg].
  \end{aligned}
\end{equation}
Here $\omega=(k^2+m_D^2)^{1/2}$. Both kernels include the vacuum response to the phase gradient, which we discuss in Section~\ref{sec:fermion_source_balance}.

At second order in gradients, we find in both models that the density contains only temporal derivatives, while the longitudinal current contains mixed temporal and spatial derivatives. For the Dirac fermion, one such term is proportional to $(\partial_z m_D)(\partial_t\alpha)$.

\subsection{Sources from the fermion field equations}
Varying the Lagrangian with respect to $\psi_L^\dagger$ gives
\begin{equation}
   -i \bar{\sigma}^\mu \partial_\mu \psi_L + m_D e^{-i\alpha} \psi_R - j_L = 0, 
\end{equation}
The resulting divergence of the left-handed current is
\begin{equation}
  \begin{aligned}
    S_L(x) &\equiv \partial_\mu j^\mu_L  \\
    &= \frac{1}{2} \left[\left\langle \partial_\mu {\psi}_L^\dagger \bar{\sigma}^\mu \psi_L \right\rangle+\left\langle {\psi}_L^\dagger \bar{\sigma}^\mu \partial_\mu \psi_L \right\rangle - \left\langle (\bar{\sigma}^\mu\partial_\mu{\psi}_L)^T (\psi_L^\dagger)^T \right\rangle-\left\langle {\psi}_L^T (\partial_\mu\psi_L^\dagger\bar{\sigma}^\mu)^T \right\rangle\right]\\
    &= \frac{1}{2} \big[i m_De^{-i\alpha}\left(\langle \psi_R^T (\psi_L^\dagger)^T\rangle- \langle \psi_L^\dagger \psi_R\rangle\right) -i \langle j_L^\dagger \psi_L\rangle -i\langle j_L^T (\psi_L^\dagger)^T\rangle + \text{h.c.}\big],
  \end{aligned}
\end{equation}
The term proportional to $m_D$ defines the mass-mixing source $S_L^m$,
and the terms involving the interaction operator $j_L$ define the
decay source $S_L^d$.

\subsubsection{Mass-mixing source}
The zeroth- and first-order mass-mixing sources vanish at the order
retained here. Their integrals are
\begin{equation}
    {}^{\text{CPV}(0)}S_L^m(x) = -m_D \operatorname{Im} \int \frac{d^4 k}{(2\pi)^4} e^{-i\alpha }\operatorname{Tr}\left( S_{RL}^{(0),<}(k,x)+S_{RL}^{(0),>}(k,x)\right)
\end{equation}
and
\begin{equation}
  \begin{aligned}
    {}^{\text{CPV}(1)}S_L^m(x)
    &= m_D^2\sum_{\sigma=0,3}(-1)^\sigma(\partial_\sigma\alpha)
       \operatorname{Re}\sum_{\lambda=<,>}\int\frac{d^4k}{(2\pi)^4}\\
    &\quad\times\operatorname{Tr}\Big[e^{-2i\alpha}\big((\partial_{k^\sigma}S^r_{RL})S^\lambda_{RL}
                         -S^\lambda_{RL}\partial_{k^\sigma}S^a_{RL}\big)\\
    &\qquad
       -\big((\partial_{k^\sigma}S^r_{RR})S^\lambda_{LL}
                   -S^\lambda_{RR}\partial_{k^\sigma}S^a_{LL}\big)
       \Big].
  \end{aligned}
\end{equation}
As in the boson case, the spatial first-order integrand is odd in $k^3$
after taking the spinor trace. The leading mass-mixing source is second
order in gradients and is given by
\begin{equation}
  \label{eq:fermionmasssource2nd}
  \begin{aligned}
    &{}^{\text{CPV}(2)}S_L^{m}(x) \\
    &=-\sum_{\sigma=0,3}\frac{1}{2}m_D \left(m_D\partial_{x^\sigma}^2\alpha + 2 \partial_{x^\sigma}m_D\partial_{x^\sigma}\alpha \right) \operatorname{Im} \sum_{\lambda = <,>}\int\frac{d^4 k}{(2\pi)^4} \operatorname{Tr}\bigg[e^{-2i\alpha}\partial_{k^\sigma}^2  S^r_{RL}  S^\lambda_{RL} \\
    &\qquad\quad {}+ e^{-2i\alpha} S^\lambda_{RL} \partial_{k^\sigma}^2 S^a_{RL}- \partial_{k^\sigma}^2 S^r_{RR} S^\lambda_{LL}  -  S^\lambda_{RR}\partial_{k^\sigma}^2 S^a_{LL}\bigg] \\
    & -\sum_{\sigma=0,3} 2m_D^2 \partial_{x^\sigma} m_D \partial_{x^\sigma}\alpha \operatorname{Re} \sum_{\lambda = <,>}\int\frac{d^4 k}{(2\pi)^4} e^{-i\alpha}\operatorname{Tr}\bigg[ \\
    &\qquad e^{-2i\alpha}\big(\partial_{k^\sigma} S^r_{RL}  S^\lambda_{RL} \partial_{k^\sigma} S^a_{RL}+ \partial_{k^\sigma} S^r_{RL} S^r_{RL} \partial_{k^\sigma} S^\lambda_{RL} + \partial_{k^\sigma} S^\lambda_{RL}  S^a_{RL} \partial_{k^\sigma} S^a_{RL}\big) \\
    &-e^{2i\alpha}\left(\partial_{k^\sigma} S^r_{RR}  S^\lambda_{LR} \partial_{k^\sigma} S^a_{LL}+\partial_{k^\sigma} S^r_{RR}  S^r_{LR}\partial_{k^\sigma} S^\lambda_{LL}+\partial_{k^\sigma} S^\lambda_{RR}  S^a_{LR} \partial_{k^\sigma} S^a_{LL}\right)\bigg]
  \end{aligned}
\end{equation}
\subsubsection{Decay source}
In position space, the decay source is
\begin{equation}
  \label{eq:fermionDsourceposition}
  \begin{aligned}
    S_L^{d}(x) &= -\lim_{y\to x}\operatorname{Im}\sum_{\lambda = <,>}\int d^4 \omega \operatorname{Tr} \left[\Pi_{LL}^{(0),r}(x,\omega) S_{LL}^\lambda(\omega,y) + \Pi_{LL}^{(0),\lambda}(x,\omega) S_{LL}^a(\omega,y)\right].
  \end{aligned}
\end{equation}
Write $F_{LL}=S^<_{LL}+S^>_{LL}$. Fourier transforming the retarded self-energy in the first term of
Eq.~\eqref{eq:fermionDsourceposition} gives
\begin{equation}
  \label{eq:fermion_decay_retarded}
  \begin{aligned}
    \left.S_L^d(x)\right|_1
      &=-\lim_{y\to x}\operatorname{Im}\operatorname{Tr}
        \left[(ia_L\bar{\sigma}^\mu\partial_{x^\mu}
               +b_L-2i\Gamma_L)F_{LL}(x,y)\right]\\
      &=a_L\partial_\mu j_L^\mu(x)-4\Gamma_L n_L(x).
  \end{aligned}
\end{equation}
The derivative acts on the first argument before taking the coincidence
limit. Hermiticity relates its real part to half the derivative of the
coincident correlator and makes the $b_L$ term vanish.
The bath KMS relation gives
$\Pi_{LL}^{(0),<}+\Pi_{LL}^{(0),>}=-4i\Gamma_L(1-2n_{F,L})$.
The second term becomes
\begin{equation}
  \label{eq:fermion_decay_gain}
  \begin{aligned}
    \left.S_L^d(x)\right|_2
      &=4\Gamma_L\operatorname{Re}
        \int d^4\omega\int\frac{d^4k}{(2\pi)^4}
        (1-2n_{F,L})e^{-ik(x-\omega)}
        \operatorname{Tr}S_{LL}^a(\omega,x)\\
      &\equiv 4\Gamma_L n_L^{\mathrm{eq}}(x).
  \end{aligned}
\end{equation}
Here $n_L^{\mathrm{eq}}$ is defined by the convolution in the gain term. In a homogeneous background it reduces to
$-\frac12\int d^4k/(2\pi)^4\,(1-2n_{F,L})\operatorname{Tr}\rho_{LL}(k)$.
Equation~\eqref{eq:fermion_decay_gain} is exact for the self-energy
ansatz above. Since $\Gamma_L=\mathcal O(g^2)$, the gain term through
$\mathcal O(g^2)$ requires only the $\mathcal O(g^0)$ part of
$n_L^{\mathrm{eq}}$ at the gradient order of interest.
With this definition, the two terms combine to give
\begin{equation}
  \label{eq:fermion_decay_relaxation}
  S_L^d(x)=a_L\partial_\mu j_L^\mu(x)
                 -4\Gamma_L\bigl(n_L(x)-n_L^{\mathrm{eq}}(x)\bigr).
\end{equation}
Moving the dispersive term to the left gives the kinetic factor
$Z_L=1-a_L$ in the continuity equation. The same factor appears in
the inverse propagator in Eq.~(B.4) of Ref.~\cite{PostmaCancel}.
For an unmixed channel at fixed bath parameters, the density relaxation
rate $4\Gamma_L/Z_L$ is twice the pole damping rate $2\Gamma_L/Z_L$.
Weldon relates the self-energy discontinuity to the relaxation rate of an occupation~\cite{Weldon}.
For the second-order calculation we retain the convolution in
Eq.~\eqref{eq:fermionDsourceposition}. The zeroth-order mixed correlator is
\begin{equation}
  \tilde{\boldsymbol{D}}^{(0)}(k,x) = -\tilde{\boldsymbol{\Pi}}^{(0)}(k,x) \tilde{\boldsymbol{S}}^{(0)}(k,x),
\end{equation}
The zeroth- and first-order decay sources are
\begin{equation}
  {}^{\text{CPV}(0)}S_L^d(x) = \operatorname{Im} \int \frac{d^4 k}{(2\pi)^4} \operatorname{Tr}\left( D_{LL}^{(0),<}(k,x)+D_{LL}^{(0),>}(k,x)\right),
\end{equation}
and
\begin{equation}
  \begin{aligned}
    {}^{\text{CPV}(1)}S_L^d(x)
    &= -m_D\sum_{\sigma=0,3}(-1)^\sigma(\partial_\sigma\alpha)
       \operatorname{Re}\sum_{\lambda=<,>}\int\frac{d^4k}{(2\pi)^4}\\
    &\quad\times\operatorname{Tr}\Big[
       e^{-i\alpha}\big((\partial_{k^\sigma}D^r_{LL})S^\lambda_{RL}
                       -D^\lambda_{LL}\partial_{k^\sigma}S^a_{RL}\big)\\
    &\qquad
       -e^{i\alpha}\big((\partial_{k^\sigma}D^r_{LR})S^\lambda_{LL}
                       -D^\lambda_{LR}\partial_{k^\sigma}S^a_{LL}\big)
       \Big].
  \end{aligned}
\end{equation}
Both vanish at the retained order. Repeating the check with the exact, unexpanded fermion propagator of Eq.~\eqref{eq:resumfermionpropagator}, we find that the first-order mass-mixing and decay sources cancel to all orders in $g$. At second order in gradients, the decay source through $\mathcal O(g^0)$ is
\begin{equation}
  \label{eq:fermiondecaysource2nd}
  \begin{aligned}
    &{}^{\text{CPV}(2)}S_L^{d}(x) \\
    &= \sum_{\sigma=0,3}\frac{1}{2} \left(m_D\partial_{x^\sigma}^2\alpha + 2 \partial_{x^\sigma} m_D \partial_{x^\sigma}\alpha \right) \operatorname{Im} \sum_{\lambda = <,>}\int\frac{d^4 k}{(2\pi)^4} \operatorname{Tr}\bigg[\partial_{k^\sigma}^2  D^r_{LL} e^{-i\alpha} S^\lambda_{RL} \\
    &\qquad\quad {}- \partial_{k^\sigma}^2 D^r_{LR} e^{i\alpha} S^\lambda_{LL} + D^\lambda_{LL} e^{-i\alpha} \partial_{k^\sigma}^2 S^a_{RL} -  D^\lambda_{LR} e^{i\alpha} \partial_{k^\sigma}^2 S^a_{LL}\bigg] \\
    & +\sum_{\sigma=0,3} 2m_D \partial_{x^\sigma}m_D \partial_{x^\sigma}\alpha \operatorname{Re} \sum_{\lambda = <,>}\int\frac{d^4 k}{(2\pi)^4} \operatorname{Tr}\bigg[ \\
    &\qquad e^{-2i\alpha}\big(\partial_{k^\sigma} D^r_{LL}  S^\lambda_{RL} \partial_{k^\sigma} S^a_{RL}+ \partial_{k^\sigma} D^r_{LL} S^r_{RL} \partial_{k^\sigma} S^\lambda_{RL} + \partial_{k^\sigma} D^\lambda_{LL}  S^a_{RL} \partial_{k^\sigma} S^a_{RL}\big) \\
    &-e^{2i\alpha}\left(\partial_{k^\sigma} D^r_{LR}  S^\lambda_{LR} \partial_{k^\sigma} S^a_{LL}+\partial_{k^\sigma} D^r_{LR}  S^r_{LR}\partial_{k^\sigma} S^\lambda_{LL}+\partial_{k^\sigma} D^\lambda_{LR}  S^a_{LR} \partial_{k^\sigma} S^a_{LL}\right)\bigg].
  \end{aligned}
\end{equation}
Appendix~\ref{app:fermion_second_order_sources} gives the evaluated
mass-mixing and decay kernels.

\subsection{Continuity equation and comparison with scalar mixing}
\label{sec:fermion_source_balance}
At $\mu_L=\mu_R=0$ and through $\mathcal O(g^0)$, the first-order
current in Eqs.~\eqref{eq:fermionfonL} and \eqref{eq:fermionfojLz}
and the second-order sources in Eqs.~\eqref{eq:fermionmasssource2nd}
and \eqref{eq:fermiondecaysource2nd} satisfy
\begin{equation}
  \partial_t\!\left({}^{\mathrm{CPV}(1)}n_L\right)
  +\partial_z\!\left({}^{\mathrm{CPV}(1)}j_L^z\right)
  ={}^{\mathrm{CPV}(2)}S_L^m+{}^{\mathrm{CPV}(2)}S_L^d .
\end{equation}
The temporal and spatial terms agree separately. The varying Dirac mass phase alone therefore induces a first-order left-handed current with a second-order divergence, even in the absence of additional number-changing interactions. In the semiclassical treatment, a varying
mass phase likewise produces a CP-violating force in the flow term of
the kinetic equation~\cite{Force}.
Our calculation treats two Weyl fields coupled by a single Dirac mass.
Majorana masses and the lepton-number violation they can induce require
a separate analysis.

The continuity equation also holds for the vacuum contribution. Even at
$T=0$, the phase gradient polarizes the fermion vacuum, and the
divergence of this current enters the mass-mixing source. Taking
$T\to0$ in Eqs.~\eqref{eq:fermionfonL} and~\eqref{eq:fermionfojLz} gives
\begin{equation}
 j_{L,\mathrm{vac}}^\mu
 = m_D^2\partial^\mu\alpha
    \int\frac{d^3 k}{(2\pi)^3}\frac{1}{2\omega^3}
 = 2m_D^2\partial^\mu\alpha
    \int\frac{d^4p_E}{(2\pi)^4}\frac{1}{(p_E^2+m_D^2)^2}.
 \label{eq:fermion_vacuum_current}
\end{equation}
The Euclidean integral is logarithmically divergent. Its coefficient
per color agrees with the soft limit of the axial--pseudoscalar
correlator in Appendix~B of Kono et al.~\cite{KonoChiral2021} after
including the pseudoscalar mass insertion $m_D\alpha$. Comelli, Pietroni
and Riotto found the same local structure, $m_D^2\partial^\mu\alpha$,
in the one-loop left-handed top current induced by a CP-violating Higgs
wall~\cite{ComelliCurrents1995,ComelliCurrents1996}. They identified the
ultraviolet pole as mixing of the fermion current with the Higgs current.

To absorb the logarithm, couple external sources $A_{L\mu}$ and
$A_{R\mu}$ to the chiral currents and add the covariant kinetic
counterterm for the complex mass background,
\begin{equation}
 \delta\mathcal L = \delta Z_M (D_\mu M)^*D^\mu M, \qquad
 D_\mu M = [\partial_\mu-i(A_{L\mu}-A_{R\mu})]M,
 \qquad M=m_De^{-i\alpha}.
\end{equation}
Differentiating with respect to $A_{L\mu}$ at zero sources gives
$2\delta Z_M m_D^2\partial^\mu\alpha$, which cancels the divergence
for an appropriate choice of $\delta Z_M$.
For the thermal response in Section~\ref{sec:numerical}, we remove the occupation-independent terms from both the current and the sources. Their separate continuity equation then gives
$\partial_\mu j_{L,\mathrm{th}}^\mu=S_{L,\mathrm{th}}^m+S_{L,\mathrm{th}}^d$.
The divergence acts on the full wall dependence of the subtracted current, including $m_D(z)$.

The boson and fermion self-energy ans\"atze treat chemical potentials
differently. For bosons the factor $k^0-\mu_I$ in the absorptive
self-energy cancels the real-axis pole of the Bose occupation in the
Wightman self-energies and enters the retarded and advanced propagators.
The Fermi occupation is finite at $k^0=\mu_I$, so we take a constant
fermion width. At fixed $a_I$, $b_I$ and $\Gamma_I$, the fermion
chemical potentials enter only the occupations. A microscopic
self-energy can itself depend on $\mu_I$.

In the plasma rest frame, the dispersive term
$\Pi^h_{LL}=a_L k_\mu\bar{\sigma}^\mu+b_L\boldsymbol{1}$ gives the
kinetic factor $Z_L=1-a_L$ in Eq.~\eqref{eq:fermion_decay_relaxation}
when $a_L\partial_\mu j_L^\mu$ is moved to the left. The
momentum-independent scalar self-energy used here gives no such term.
With constant $a_I$ and $b_I$, the fermion propagator has only branches
connected to free particle and antiparticle poles. An energy-dependent
self-energy can also support a collective hole branch, as in the
fermionic hard-thermal-loop example in
Appendix~\ref{app:fermion_ctp_coefficients}.

In both sectors the leading number density depends on the ratio of the
two widths, whereas the leading spatial current is independent of both
widths [Eqs.~\eqref{eq:fonL}, \eqref{eq:fojLz}, \eqref{eq:fermionfonL}
and \eqref{eq:fermionfojLz}]. Exchanging $L$ and $R$ reverses both the
mass phase $\alpha$ and the width difference $\Gamma_L-\Gamma_R$.
The terms proportional to $(\partial_t\alpha)(\Gamma_L-\Gamma_R)$
therefore add in $n_L^{\mathrm{CPV}}+n_R^{\mathrm{CPV}}$, which vanishes
for equal widths but can be nonzero for unequal widths. At this order,
the time derivative of the summed density thus comes entirely from
$S_L^d+S_R^d$. With prescribed bath self-energies, these equations track the charge of the mixing fields. Testing conservation of the total charge requires the bath current as well.

\section{Charge transport and transfer to other species}
\label{sec:transfer}
For both systems the phase gradient induces a longitudinal flux at first order, Eqs.~\eqref{eq:fojLz} and~\eqref{eq:fermionfojLz}. Together with the densities in Eqs.~\eqref{eq:fonL} and~\eqref{eq:fermionfonL}, these fluxes account for the sum of the mass-mixing and decay sources through the continuity equation at $\mu_L=\mu_R=0$. In fluid treatments, CP-conserving reactions redistribute charge in response to local chemical imbalances~\cite{CiriglianoYukawa}. Here the varying mass phase directly determines the current of the mixing fields. Two further questions concern the distance over which this charge propagates and its transfer to other species.

The local gradient expansion also applies in the tails of a smooth wall wherever the phase still varies and the expansion remains controlled. Higher gradients can correct the density there. In a region where the background is exactly constant, every local derivative vanishes. Determining whether a density persists there requires retaining propagation between the wall and the observation point in the CTP integrals, together with the appropriate state and boundary conditions.

Charge transfer can be studied through the loop-corrected lesser propagator of a species whose mass is independent of $\alpha$. At first order in gradients, the varying part of the mass is inserted on each line that carries it, one line at a time, including the mass vertex and its Hermitian conjugate. The phase gradient on that line supplies the CP violation even when the interaction between species conserves CP.

\begin{figure}[tp]
  \centering
  \begin{minipage}[b]{0.48\linewidth}
  \centering
  \resizebox{\linewidth}{!}{%
  \begin{tikzpicture}[
    line width=0.9pt,
    every node/.style={font=\large},
    fermion/.style={
        postaction={decorate},
        decoration={markings,
            mark=at position 0.5 with {\arrow{Triangle}}
        }
    },
    boson/.style={dash pattern=on 3pt off 2.5pt}
]
    \coordinate (L) at (-4.0,0);
    \coordinate (A) at (-1.5,0);
    \coordinate (B) at ( 1.5,0);
    \coordinate (R) at ( 4.0,0);
    \coordinate (M) at ( 0,1.5);

    \draw[fermion] (L) -- (A);
    \draw[fermion] (A) -- (B);
    \draw[fermion] (B) -- (R);

    \draw[boson] (A) arc[start angle=180, end angle=90, radius=1.5cm];
    \draw[boson] (B) arc[start angle=0,   end angle=90, radius=1.5cm];

\foreach \point in {A,B} {
    \fill (\point) circle[radius=2.5pt];
}

\begin{scope}[shift={(M)}]
    \draw (-4pt,-4pt) -- (4pt,4pt);
    \draw (-4pt,4pt) -- (4pt,-4pt);
\end{scope}

    \node[above=3pt] at (-2.75,0) {$S_{11} $};       
    \node[below=3pt] at ( 0,0) {$ S_{RR}$};       
    \node[above=3pt] at ( 2.75,0) {$ S_{11}$};       
    \node[above left]  at (-1.06,1.06) {$G_{LL} $}; 
    \node[above right] at ( 1.06,1.06) {$G_{RL} $}; 
\end{tikzpicture}
  }%
  \par\smallskip
  (a)
\end{minipage}\hfill
\begin{minipage}[b]{0.48\linewidth}
  \centering
  \resizebox{\linewidth}{!}{%
  \begin{tikzpicture}[
    line width=0.9pt,
    every node/.style={font=\large},
    fermion/.style={
        postaction={decorate},
        decoration={markings,
            mark=at position 0.5 with {\arrow{Triangle}}
        }
    },
    boson/.style={dash pattern=on 3pt off 2.5pt}
]
    \coordinate (L) at (-4.0,0);
    \coordinate (A) at (-1.5,0);
    \coordinate (B) at ( 1.5,0);
    \coordinate (R) at ( 4.0,0);
    \coordinate (M) at ( 0,1.5);

    \draw[fermion] (L) -- (A);
    \draw[fermion] (A) -- (B);
    \draw[fermion] (B) -- (R);

    \draw[boson] (A) arc[start angle=180, end angle=90, radius=1.5cm];
    \draw[boson] (B) arc[start angle=0,   end angle=90, radius=1.5cm];

\foreach \point in {A,B} {
    \fill (\point) circle[radius=2.5pt];
}

\begin{scope}[shift={(M)}]
    \draw (-4pt,-4pt) -- (4pt,4pt);
    \draw (-4pt,4pt) -- (4pt,-4pt);
\end{scope}

    \node[above=3pt] at (-2.75,0) {$S_{11} $};       
    \node[below=3pt] at ( 0,0) {$ S_{RR}$};       
    \node[above=3pt] at ( 2.75,0) {$ S_{11}$};       
    \node[above left]  at (-1.06,1.06) {$G_{LR} $}; 
    \node[above right] at ( 1.06,1.06) {$G_{LL} $}; 
\end{tikzpicture}
  }%
  \par\smallskip
  (b)
\end{minipage}
\caption{One-loop Yukawa contributions to the $\psi_1$ two-point function. The cross denotes a gradient insertion on the scalar line.}
\label{fig:yukawa}
\end{figure}
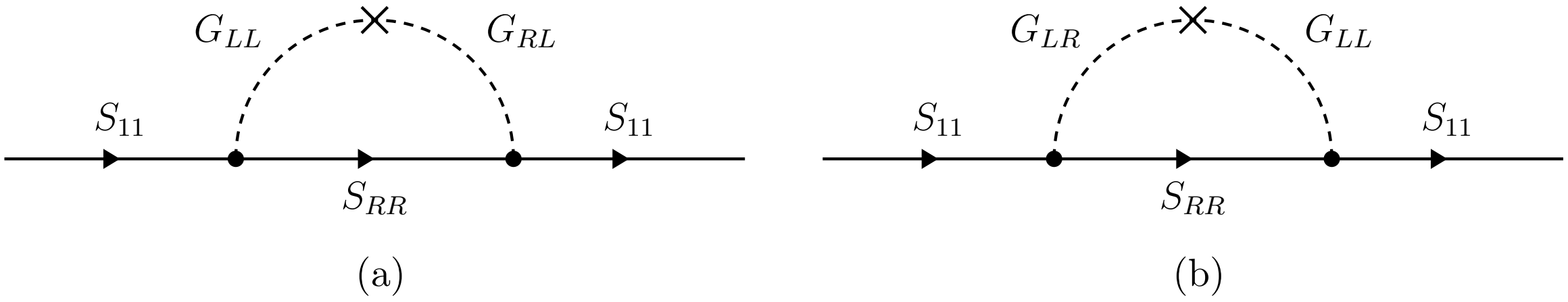

Consider the Yukawa interaction
\begin{equation}
  \mathcal L_Y = \lambda\,\Phi_L\,\psi_1^\dagger\psi_R + \text{h.c.},
\end{equation}
where $\psi_1$ is a left-handed field distinct from $\psi_L$, and $L$ on $\Phi_L$ labels scalar flavor. We assume that $\psi_1$ and $\psi_R$ have no mass mixing and that $\langle\Phi_L\rangle=0$, so the Yukawa interaction generates no mass term between them. The one-loop correction to the $\psi_1$ propagator contains an internal $\Phi_L$ line and an internal $\psi_R$ line. When the scalar mixing mass varies, its gradient enters the scalar propagator, as shown in Figure~\ref{fig:yukawa}. The two panels correspond to the mass insertion and its Hermitian conjugate.

\begin{figure}[tp]
  \centering
  \begin{minipage}[b]{0.48\linewidth}
  \centering
  \resizebox{\linewidth}{!}{%
  \begin{tikzpicture}[
    line width=0.9pt,
    every node/.style={font=\large},
    fermion/.style={
        postaction={decorate},
        decoration={markings,
            mark=at position 0.5 with {\arrow{Triangle}}
        }
    },
    boson/.style={dash pattern=on 3pt off 2.5pt}
]
    \coordinate (L) at (-4.0,0);
    \coordinate (A) at (-1.5,0);
    \coordinate (B) at ( 1.5,0);
    \coordinate (R) at ( 4.0,0);
    \coordinate (M) at ( 0,1.5);

    \draw[fermion] (L) -- (A);
    \draw[boson] (A) -- (B);
    \draw[fermion] (B) -- (R);

    \draw[fermion] (A) arc[start angle=180, end angle=90, radius=1.5cm];
    \draw[fermion] (M) arc[start angle=90, end angle=0, radius=1.5cm];

\foreach \point in {A,B} {
    \fill (\point) circle[radius=2.5pt];
}

\begin{scope}[shift={(M)}]
    \draw (-4pt,-4pt) -- (4pt,4pt);
    \draw (-4pt,4pt) -- (4pt,-4pt);
\end{scope}

    \node[above=3pt] at (-2.75,0) {$S_{11} $};       
    \node[below=3pt] at ( 0,0) {$ G_{LL}$};       
    \node[above=3pt] at ( 2.75,0) {$ S_{11}$};       
    \node[above left]  at (-1.06,1.06) {$S_{RR} $}; 
    \node[above right] at ( 1.06,1.06) {$S_{LR} $}; 
\end{tikzpicture}
  }%
  \par\smallskip
  (a)
\end{minipage}\hfill
\begin{minipage}[b]{0.48\linewidth}
  \centering
  \resizebox{\linewidth}{!}{%
  \begin{tikzpicture}[
    line width=0.9pt,
    every node/.style={font=\large},
    fermion/.style={
        postaction={decorate},
        decoration={markings,
            mark=at position 0.5 with {\arrow{Triangle}}
        }
    },
    boson/.style={dash pattern=on 3pt off 2.5pt}
]
    \coordinate (L) at (-4.0,0);
    \coordinate (A) at (-1.5,0);
    \coordinate (B) at ( 1.5,0);
    \coordinate (R) at ( 4.0,0);
    \coordinate (M) at ( 0,1.5);

    \draw[fermion] (L) -- (A);
    \draw[boson] (A) -- (B);
    \draw[fermion] (B) -- (R);

    \draw[fermion] (A) arc[start angle=180, end angle=90, radius=1.5cm];
    \draw[fermion] (M) arc[start angle=90, end angle=0, radius=1.5cm];

\foreach \point in {A,B} {
    \fill (\point) circle[radius=2.5pt];
}

\begin{scope}[shift={(M)}]
    \draw (-4pt,-4pt) -- (4pt,4pt);
    \draw (-4pt,4pt) -- (4pt,-4pt);
\end{scope}

    \node[above=3pt] at (-2.75,0) {$S_{11} $};       
    \node[below=3pt] at ( 0,0) {$ G_{LL}$};       
    \node[above=3pt] at ( 2.75,0) {$ S_{11}$};       
    \node[above left]  at (-1.06,1.06) {$S_{RL} $}; 
    \node[above right] at ( 1.06,1.06) {$S_{RR} $}; 
\end{tikzpicture}
  }%
  \par\smallskip
  (b)
\end{minipage}
\caption{One-loop Yukawa contributions to the $\psi_1$ two-point function. The cross denotes an insertion of the varying Dirac mass on the internal fermion line.}
\label{fig:fermionyukawa}
\end{figure}
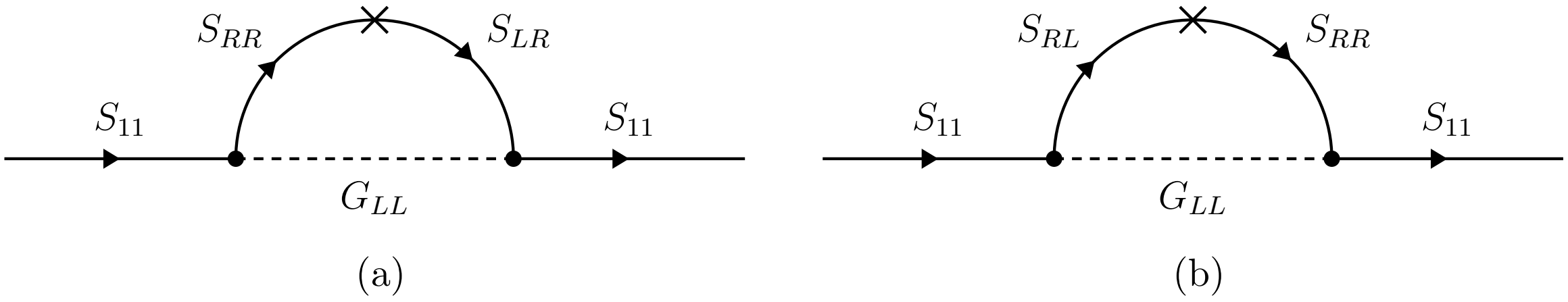

When the Dirac mass coupling $\psi_L$ to $\psi_R$ varies, the gradient instead enters the internal fermion propagator, as shown in Figure~\ref{fig:fermionyukawa}. In each case, the sum of the two diagrams gives a one-loop correction to the $\psi_1$ propagator proportional to $|\lambda|^2$. Its lesser component determines the induced $\psi_1$ current at vanishing bath chemical potentials. We leave the evaluation of the thermal loop integrals to future work.

\FloatBarrier
\section{Numerical comparison with the VEV insertion approximation}
\label{sec:numerical}

For mixing stops and chiral top quarks, we ask how changes in the CP-violating source at the wall center are reflected in the charge density across the wall. We vary the masses and thermal widths at fixed wall profile and compare the local density carried by the resummed current with the density obtained from the VIA diffusion equation.

\subsection{Wall benchmark and observables}

We use the plasma-frame coordinate $z=z_{\rm lab}+v_w t$, with the symmetric phase at $z<0$. The stop and top benchmarks share $T=100~\mathrm{GeV}$, $L_wT=50$ and $v_w=0.10$. We write the kink profile~\cite{JohnSchmidtWall} as
\begin{equation}
  h_u(z)\equiv\langle H_u^0(z)\rangle=1.5\,T f(z), \qquad \alpha(z)=0.15\,f(z), \qquad
  f(z)=\frac{1+\tanh(3z/L_w)}{2}.
  \label{eq:numerical_wall_profile}
\end{equation}
We specify the relative phase profile independently of the Higgs amplitude. For mixing stops, $v^2(z)=|y_t A_t|h_u(z)$ with $y_t=1$, $A_t=400~\mathrm{GeV}$ and $M_R=800~\mathrm{GeV}$. We include the common thermal correction $(60.8~\mathrm{GeV})^2$ in the two diagonal stop mass inputs before diagonalization. For chiral tops we vary the amplitude of $m_D(z)\propto h_u(z)$, labeled by $m_D(0)$. For the comparison with VIA we set $a_L=a_R=0$ and $b_L=b_R=60.8~\mathrm{GeV}$. The VIA chiral kernel uses the thermal particle and hole poles~\cite[Appendix~B]{Cline}. We solve their dispersion equations numerically and match their common zero-momentum energy to that of the unmixed resummed modes. At finite momentum the VIA kernel resolves the two thermal branches, while the resummed benchmark uses the simpler chiral self-energy of Section~\ref{sec:dirac}. Differences between the fermion curves can therefore reflect the choice of thermal spectrum as well as the treatment of mixing. In both sectors $\gamma_I$ is the physical pole width, with $\gamma_I=2\Gamma_I$ for the fermions in this benchmark. We sum over three colors.

In Figures~\ref{fig:stops_mass}--\ref{fig:tops_width}, the resummed source is the wall-center divergence $S_L^m+S_L^d=\partial_z(j_L^z+v_w n_L)$, evaluated with the analytic narrow-width kernels of Appendices~\ref{app:boson_second_order_sources} and~\ref{app:fermion_second_order_sources}. The density profiles follow from the corresponding first-order currents in Eqs.~\eqref{eq:fonL} and~\eqref{eq:fermionfonL}. The current vanishes far from the wall on both sides, so its divergence integrates to zero even when the wall-center source is large.

VIA treats the mixing mass as an insertion in the unmixed thermal propagators. At first order in gradients the CP-violating source is proportional to $(v^2)^2\partial_t\alpha$ for stops and to $m_D^2\partial_t\alpha$ for tops~\cite{Lee,PostmaNLO}. The VIA densities follow from a diffusion equation with this source and relaxation due to mass mixing. We decouple spectator species and use $D_L=D_R=D=6/T$. The $L$ and $R$ sources and collision terms are opposite, so the summed diffusion equation and the vanishing boundary densities imply $n_R=-n_L$. The left equation is
\begin{equation}
 D n_L''-v_w n_L'-
 \left[\Gamma_-(z)\!\left(\frac{1}{k_L}+\frac{1}{k_R}\right)
      +\Gamma_+(z)\!\left(\frac{1}{k_L}-\frac{1}{k_R}\right)\right]n_L
 =-S_L^{\rm VIA}(z).
 \label{eq:benchmark_via_diffusion}
\end{equation}
Primes denote $z$ derivatives, $k_I$ are single-component statistical weights, and $\Gamma_\pm(z)$ are the relaxation rates due to mass mixing, including the conversion from chemical potentials to densities. For equal statistical weights the $\Gamma_+$ term cancels. We solve with the smooth wall profile for both the source and the rates. We check that the VIA profiles shown are stable when the numerical boundaries are moved farther from the wall.

The sign of a source in the comparison therefore has a different interpretation in the two calculations. Opposite source signs can accompany densities of the same sign. The resummed divergence measures the slope of $j_L^z+v_w n_L$ and can change sign while $n_L$ retains its sign. For a positive effective relaxation rate, a VIA source of fixed sign induces a density of the same sign. Its extent also depends on diffusion. Far into the symmetric phase the source and relaxation rate vanish. Equation~\eqref{eq:benchmark_via_diffusion} then gives $n_L\propto\exp(v_w z/D)$, with diffusion length $D/v_w$. The resummed density measures the local response to the wall gradients. A comparison of the charge transported far ahead of the wall would also require its propagation from the wall, as discussed in Section~\ref{sec:transfer}.

\subsection{Mixing stops}
\paragraph{Resonance at degenerate stop masses.}
At $M_L=M_R$ the common thermal correction makes $\omega_L(k)=\omega_R(k)$ at every momentum. In the VIA source, the difference of the Bose occupations cancels one power of the retarded--advanced energy denominator. Up to smooth thermal and kinematic weights, the resonant part of the momentum kernel is proportional to~\cite{Lee,PostmaNLO}
\begin{equation}
 (v^2)^2\,\partial_t\alpha\,
 \frac{\gamma_L+\gamma_R}
 {(\omega_L-\omega_R)^2+(\gamma_L+\gamma_R)^2}.
 \label{eq:stop_via_resonant_factor}
\end{equation}
As a function of the energy difference $\omega_L-\omega_R$, the resonance has a half-width at half-maximum of $\gamma_L+\gamma_R$. Smaller thermal widths give a higher, narrower peak. For $\gamma_L=\gamma_R=\gamma\to0^+$, the source vanishes at fixed $\omega_L\ne\omega_R$ and grows as $1/\gamma$ when $\omega_L=\omega_R$. For degenerate unmixed stop masses, the energies are equal at every momentum, so the enhancement survives momentum integration.

Resumming $v^2$ in the propagator gives an avoided crossing. The resummed response depends on the energy splitting and the mixing angle through
\begin{equation}
 \omega_1-\omega_2=
 \frac{\sqrt{(M_L^2-M_R^2)^2+4v^4}}{\omega_1+\omega_2},
 \qquad
 c_\theta^2s_\theta^2=
 \frac{v^4}{(M_L^2-M_R^2)^2+4v^4}.
 \label{eq:stop_resummed_mixing_weight}
\end{equation}
For $v^2\ne0$ the mass eigenstates have different energies at every finite momentum, even at zero width. Mixing is maximal at $M_L=M_R$, where $c_\theta^2s_\theta^2=1/4$, and remains large for $|M_L^2-M_R^2|\lesssim 2v^2$. The mass splitting removes the pinch singularity in the zero-width limit. At $M_L=M_R$, the energy splitting is largest at zero momentum and decreases as the momentum increases. Section~\ref{sec:source_accuracy} examines how finite widths affect the leading source.

\begin{figure}[tp]
 \centering\includegraphics[width=\linewidth]{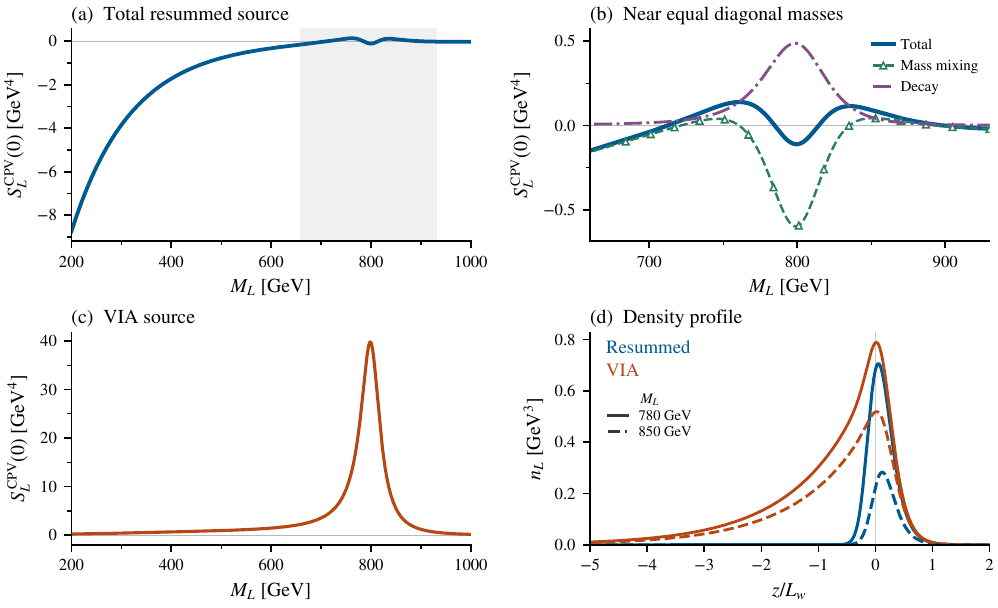}
 \caption{Mixing stops at $M_R=800~\mathrm{GeV}$, $A_t=400~\mathrm{GeV}$ and $\gamma_L=\gamma_R=10~\mathrm{GeV}$. Panels~(a) and~(c) give the leading resummed and VIA sources at $z=0$ for $200\leq M_L\leq1000~\mathrm{GeV}$. Panel~(b) shows the leading mass-mixing and decay contributions and their sum in the shaded interval. Panel~(d) compares the leading resummed and VIA density profiles for $M_L=780$ and $850~\mathrm{GeV}$ on the same scale.}
 \label{fig:stops_mass}
\end{figure}

At leading order in the narrow-width expansion and for equal widths, our thermal stop density and spatial current agree with the results of Carena et al.~\cite{CarenaCurrents} after changing the sign before $3/(m_1^2-m_2^2)$ from minus to plus in their Eq.~(2.18) for $\mathcal G_B$.

Figure~\ref{fig:stops_mass}(b) resolves the structure near equal diagonal masses. As $M_L$ approaches $M_R$ from either side, the mass-mixing source changes sign and develops a negative minimum, while the decay source peaks with the opposite sign. Their temporal parts nearly cancel at degeneracy. The remaining negative spatial divergence sets the depth of the central dip. On either side the temporal divergence becomes positive and exceeds the spatial contribution in magnitude, producing the two shoulders. The VIA source in panel~(c) instead has a single peak, set by the degeneracy of the unmixed energies in Eq.~\eqref{eq:stop_via_resonant_factor}.

The origin of the two shoulders can be read from the density. For equal widths and an energy splitting small on the thermal scale, Eq.~\eqref{eq:fonL} gives $n_L$ proportional to $c_\theta^2s_\theta^2\,\partial_t\alpha$ times a smooth Bose susceptibility. At $M_L=M_R$ the mixing factor is $1/4$ throughout the wall interior. The density therefore peaks near the wall center, where $\alpha''(0)=0$, so $\partial_t n_L=v_w\partial_z n_L$ is small even though the density is large. Away from degeneracy the mixing factor grows as $v^2(z)$ increases across the wall. At fixed $z$, its wall derivative as a function of $M_L$ is largest at $|M_L^2-M_R^2|=2v^2(z)$, by Eq.~\eqref{eq:stop_resummed_mixing_weight}. This gives a positive contribution to $\partial_t n_L$ on both sides of degeneracy and accounts for the locations of the shoulders up to the mass dependence of the susceptibility and flux.

Farther down the low-mass branch in panel~(a), the lighter stop has a larger Bose occupation. Here $c_\theta^2s_\theta^2\simeq v^4/(M_R^2-M_L^2)^2$ is small but grows across the wall. Since $\alpha''(0)=0$, differentiating this mixing factor gives the dominant term in $\partial_z j_L^z(0)$ in Eq.~\eqref{eq:fojLz}. The resulting negative flux divergence grows in magnitude toward lower $M_L$ and dominates the temporal divergence, which carries two powers of $v_w$.

\begin{figure}[tp]
 \centering\includegraphics[width=\linewidth]{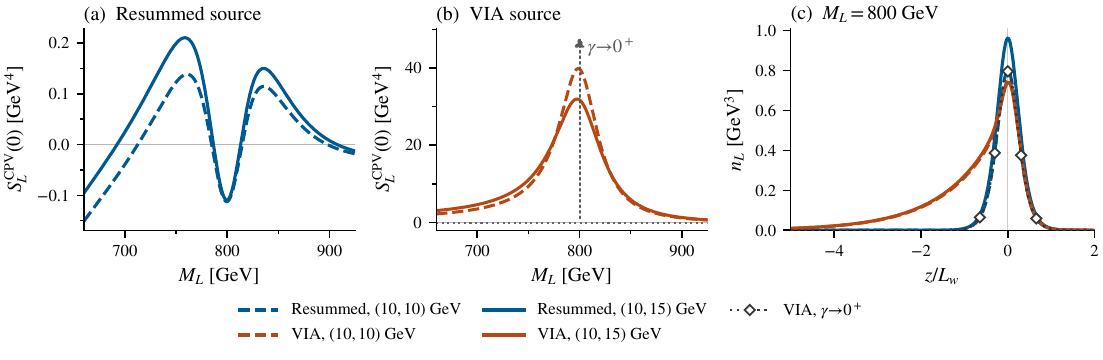}
 \caption{Width dependence for mixing stops at $M_R=800~\mathrm{GeV}$. Dashed and solid curves give $(\gamma_L,\gamma_R)=(10,10)$ and $(10,15)~\mathrm{GeV}$, respectively. Panels~(a) and~(b) show the resummed and VIA wall-center sources. The dotted baseline and arrow in panel~(b) give the VIA source as $\gamma_L=\gamma_R=\gamma\to0^+$. In panel~(c) the dotted curve with open diamonds gives the finite VIA density in the same limit at $M_L=M_R$. The dashed resummed curves also give their common-width limits.}
 \label{fig:stops_width}
\end{figure}

The large source on the low-mass branch need not imply a large density. The density depends on the mixing factor itself, and the enhanced Bose occupation only partly compensates for the loss of mixing far from degeneracy. For both masses in panel~(d) the resummed density peak is lower and lies farther into the broken phase than the equal-mass peak in Figure~\ref{fig:stops_width}(c). The mass closer to degeneracy gives the larger density. Both VIA profiles have the same sign as the local densities and extend farther into the symmetric phase. At the low-mass end of the source scan, the effective VIA relaxation rate becomes negative in the broken phase. Equation~\eqref{eq:benchmark_via_diffusion} then has no homogeneous solution that decays as $z\to+\infty$. Both masses used for the density comparison admit a decaying solution. A sign change of the VIA relaxation rate far from degeneracy was also observed in Ref.~\cite{AkulaNMSSM}.

\paragraph{Narrow-width limit and density.}
Figure~\ref{fig:stops_width} shows why the VIA source is resonantly enhanced while the density remains finite. The sum $\gamma_L+\gamma_R$ sets the width of the VIA resonance in Eq.~\eqref{eq:stop_via_resonant_factor}. Away from equal diagonal masses the VIA source tends to zero with the width. At $M_L=M_R$, the leading resonant terms in the stop source and relaxation rate are both proportional to $1/\gamma$ as $\gamma_L=\gamma_R=\gamma\to0$. The arrow in panel~(b) marks this formal limit of the local VIA source. At fixed wall thickness, taking the width to zero eventually takes the calculation outside the local gradient expansion. An analogous factor appears in Kainulainen's fermionic VIA derivation, where the same singular integral multiplies the CP-violating source and the CP-even relaxation term~\cite{KainulainenThermal}. At fixed position inside the wall, the resonant relaxation term dominates the derivatives in Eq.~\eqref{eq:benchmark_via_diffusion} as $\gamma\to0^+$. The ratio of the source to the effective relaxation rate then gives $n_L^{\rm VIA}\to k_LT^2\partial_t\alpha/12$. This is the diamond-marked profile in panel~(c). It peaks at the wall center, where the phase gradient is largest. Taking the equal-width limit at fixed nonzero mixing, then expanding Eq.~\eqref{eq:fonL} for a splitting small on the thermal scale, gives the same susceptibility at maximal mixing. As the diagonal masses separate, both the VIA source and the relaxation rate decrease. Their ratio controls the density only while relaxation dominates the diffusion equation.

Figure~\ref{fig:stops_width} illustrates two distinct width dependences. In VIA the sum $\gamma_L+\gamma_R$ sets the width of the resonance. Raising $\gamma_R$ at fixed $\gamma_L$ broadens the peak in panel~(b) and lowers its height. The two VIA density profiles in panel~(c) remain close because the same resonant factor enters the source and the relaxation rate. In the resummed calculation the density depends on the ratio of the widths. At maximal mixing and small energy splitting, Eq.~\eqref{eq:fonL} gives $n_L\propto 2\gamma_R/(\gamma_L+\gamma_R)$, so increasing $\gamma_R$ enhances the density peak. The total source at degeneracy changes little because the temporal derivative of the density is small at the wall center, and the spatial flux is independent of the widths at this order. Away from degeneracy the shoulders rise appreciably [panel~(a)]. For both width choices the VIA density extends into the symmetric phase while the resummed density remains localized near the wall.

\subsection{Chiral top quarks}
\begin{figure}[tp]
 \centering\includegraphics[width=\linewidth]{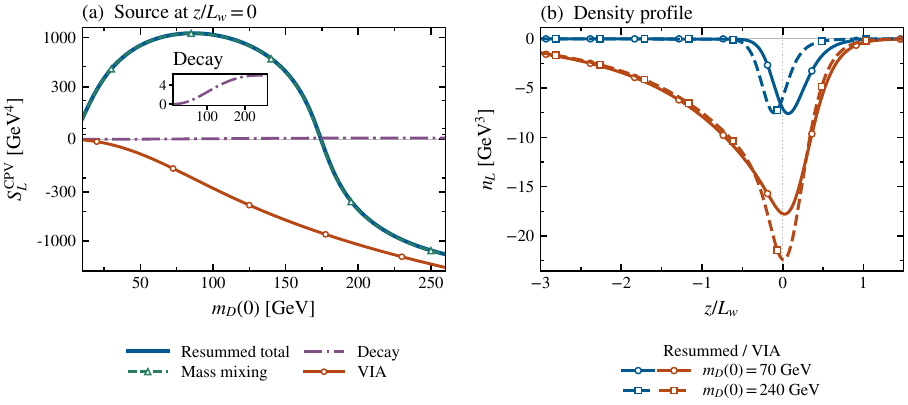}
 \caption{Chiral tops with $\gamma_L=\gamma_R=10~\mathrm{GeV}$. Panel~(a) gives the signed wall-center sources on an inverse hyperbolic sine scale. The inset resolves the decay contribution in the same units. Panel~(b) compares the leading resummed and VIA density profiles for $m_D(0)=70$ and $240~\mathrm{GeV}$.}
 \label{fig:tops_mass}
\end{figure}

\paragraph{Particle and hole contributions to the source.}
The VIA helicity projection pairs the particle pole of one chirality with the hole pole of the other in the term that can become resonant~\cite[Appendix~B]{Cline}. The energy difference is $\omega_{p,L}(k)-\omega_{h,R}(k)$ or the same difference with $L$ and $R$ exchanged. Equal thermal masses make these energies equal only at $k=0$. As the widths shrink, the radial measure $k^2\,dk$ suppresses the region where the two poles approach one another. Consequently the integrated VIA source tends to zero as $\gamma_L=\gamma_R\to0$. Scanning $m_D(0)$ at fixed thermal masses and widths leaves the particle and hole energies fixed because the Dirac mass enters the VIA kernel only as the insertion factor $m_D^2\partial_t\alpha$. This accounts for the quadratic source in Figure~\ref{fig:tops_mass}(a).

The leading resummed kernels depend on the Dirac mass through $\omega=\sqrt{k^2+m_D^2}$. The common shift $b$ drops out at this order. Removing the occupation-independent term and integrating Eq.~\eqref{eq:fermionfojLz} by parts in radial momentum gives the thermal flux summed over colors,
\begin{equation}
 j_{L,\mathrm{th}}^z=\frac{3m_D^2\,\partial_z\alpha}{2\pi^2}
   \int_0^\infty dk\,\frac{n_F(\omega)}{\omega}.
 \label{eq:numerical_top_flux}
\end{equation}
At the wall center $\alpha''(0)=0$, so the spatial divergence in Figure~\ref{fig:tops_mass}(a) probes how the flux in Eq.~\eqref{eq:numerical_top_flux} changes with the Dirac mass. At small mass the explicit $m_D^2$ factor drives the rise of the source. As the Fermi occupation falls, the flux gains less per fractional increase in mass and the source turns down. Once the flux passes its maximum, its spatial divergence changes sign. The temporal contribution is suppressed by $v_w^2$ and shifts the zero of the total source only slightly.

For equal widths the thermal part of Eq.~\eqref{eq:fermionfonL} gives $n_L\propto -m_D^2\,\partial_t\alpha\int d^3k\,n_F(\omega)/\omega^3$. This accounts for the negative resummed profiles in panel~(b) and their rapid decay on the symmetric side, where both the mass and the phase gradient vanish. In VIA, the source and the relaxation rate both grow as $m_D^2$. The density reflects their competition, and diffusion carries it ahead of the wall.

\begin{figure}[tp]
 \centering\includegraphics[width=\linewidth]{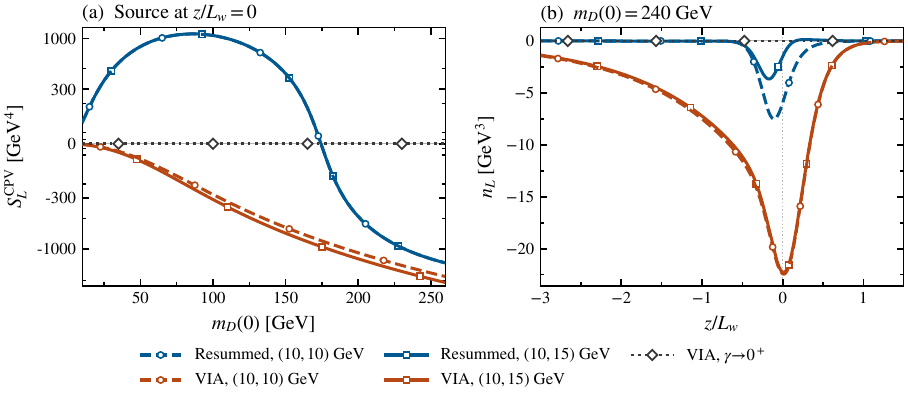}
 \caption{Width dependence for chiral tops at $\gamma_L=10~\mathrm{GeV}$. Dashed lines with circles give $\gamma_R=10~\mathrm{GeV}$ and solid lines with squares give $\gamma_R=15~\mathrm{GeV}$. Panel~(a) compares the wall-center sources on the scale of Figure~\ref{fig:tops_mass}(a). Panel~(b) gives the density profiles at $m_D(0)=240~\mathrm{GeV}$. Dotted lines with open diamonds mark the common zero-width VIA limit in both panels.}
 \label{fig:tops_width}
\end{figure}

\paragraph{Width dependence.}
Raising $\gamma_R$ at fixed $\gamma_L$ leaves the resummed source curves nearly coincident in Figure~\ref{fig:tops_width}(a), but makes the density minimum appreciably shallower in panel~(b). The longitudinal flux, which dominates the wall-center divergence, is independent of the widths at this order. The width-dependent term in Eq.~\eqref{eq:fermionfonL} is proportional to $\Gamma_L-\Gamma_R$. It vanishes for equal widths and partially cancels the negative thermal density for the unequal widths shown. In VIA, finite widths allow the particle and hole peaks to overlap over a range of momenta. For the larger $\gamma_R$ the VIA source is more negative over the mass scan. Its density profile changes little, since the width also enters the relaxation rate. Both profiles have a diffusion tail. Along the path $\gamma_L=\gamma_R=\gamma\to0^+$, the particle--hole overlap shrinks to zero phase space. The VIA source and induced density both vanish, as shown by the horizontal diamond-marked lines in panels~(a) and~(b). Along this same path the analytic resummed curves coincide with the finite equal-width curves.

In the symmetric tail the Dirac mass tends to zero. For the equal thermal shifts used here, the particle and hole poles still meet in real energy at zero momentum [Appendix~\ref{app:fermion_ctp_coefficients}].

\subsection{Accuracy of the expansion in the bath coupling}
\label{sec:source_accuracy}
\begin{figure}[tp]
 \centering\includegraphics[width=\linewidth]{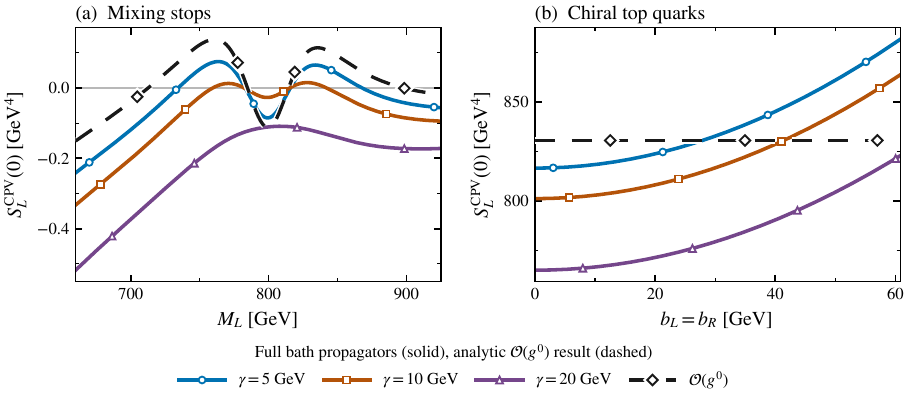}
 \caption{Thermal wall-center source from the full bath propagators (solid curves) and the analytic $\mathcal O(g^0)$ kernels (black dashes with open diamonds). The legend gives the common pole width $\gamma_L=\gamma_R=\gamma$. Panel~(a) varies $M_L$ across the resonance region at $M_R=800~\mathrm{GeV}$ and $A_t=400~\mathrm{GeV}$. Panel~(b) varies the common shift $b_L=b_R$ at fixed $m_D(0)=50~\mathrm{GeV}$ and $a_L=a_R=0$.}
 \label{fig:source_accuracy}
\end{figure}

At the same order in wall gradients, Figure~\ref{fig:source_accuracy} compares the thermal wall-center source obtained from the analytic $\mathcal O(g^0)$ kernels (dashed) with the result obtained from the full bath propagators (solid). The analytic kernels are independent of a common pole width at this order. For the solid curves, we evaluate the four-momentum integrals for $n_L$ and $j_L^z$ in Eqs.~\eqref{eq:fonL} and~\eqref{eq:fojLz} for stops and Eqs.~\eqref{eq:fermionfonL} and~\eqref{eq:fermionfojLz} for tops, keeping the bath self-energies unexpanded. The source then follows from $S_L(0)=\partial_z(j_L^z+v_w n_L)|_{z=0}$. Both calculations omit the occupation-independent terms and keep the source through second order in wall gradients, even for the narrowest poles where the locality condition is most restrictive.

In the stop scan [Figure~\ref{fig:source_accuracy}(a)], the narrowest finite-width curve resolves the central dip and both positive shoulders. As the common width increases, the shoulders lose height and the dip becomes less distinct. For the widest poles the positive shoulders have disappeared, even though the source at $M_L=M_R$ happens to lie close to the analytic value. Agreement at the center alone would therefore miss the change of shape. Near equal diagonal masses the physical energies remain split by the mixing, but their separation decreases with momentum. The analytic stop kernel assumes well-separated poles, $\gamma_1+\gamma_2\ll|\omega_1-\omega_2|$, over the momenta that carry the source. Once the two quasiparticle peaks overlap appreciably, the unexpanded propagators smooth out this structure. The same condition applies locally to the density profiles. At equal diagonal masses [Figure~\ref{fig:stops_width}(c)], the splitting decreases toward the symmetric phase as the Higgs background vanishes. Evaluating the density with the unexpanded propagators there gives a smaller symmetric-side tail than the analytic kernel.

For tops with $a_L=a_R=0$, $b_L=b_R=b$ and $\gamma_L=\gamma_R=\gamma$, the two retarded poles at fixed helicity lie at $k^0=b\pm\omega-i\gamma$, where $\omega=\sqrt{k^2+m_D^2}$ [Eq.~\eqref{eq:fermion_compact_poles}]. The common shift moves both poles equally, leaving their real-energy separation at $2\omega$. The poles are well separated when $\gamma_L+\gamma_R\ll 2\omega$, or $\gamma\ll\omega$, for the momenta that carry the source. In Figure~\ref{fig:source_accuracy}(b) the analytic curve is horizontal because the common thermal shift drops out at $\mathcal O(g^0)$. In the full bath propagators, the shift changes the occupations sampled by the poles, and the source rises across the scan. Increasing the common width lowers the curve at a given shift. The two narrower curves cross the analytic result, with the crossing moving to larger thermal shift as the width increases. The widest curve stays below it throughout the interval shown.

\section{Conclusion}
\label{sec:conclusion}

A varying mass phase $\alpha$ across a bubble wall induces a CP-odd current even at vanishing bath chemical potentials. For two mixing complex scalars and for chiral fermions coupled by a complex Dirac mass, we obtained the current by resumming the local mass matrix before expanding in wall gradients, retaining the full dependence on thermal mass shifts and damping widths within our bath model. Our power counting $L_w^{-1}\ll\gamma_i\ll T$ combines the locality condition with the narrow-width assumption. At first order the density is proportional to $\partial_t\alpha$ and the longitudinal flux to $\partial_z\alpha$. At leading order in the narrow-width expansion, the charge density can depend on the ratio $\Gamma_L/\Gamma_R$, while the longitudinal flux is independent of both widths. At second order in gradients the independently calculated mass-mixing and decay sources sum to $\partial_t n_L+\partial_z j_L^z$, with the temporal and spatial contributions agreeing separately. The resummed two-point functions thus give a local continuity equation for the induced charge without a chemical-potential ansatz. In the collisionless scalar example of Appendix~\ref{app:mass_basis_transport}, a varying phase induces off-diagonal coherence at first order in gradients, while the mass-mixing source vanishes at that order.

At equal diagonal boson masses, the resonant factors in the VIA source and relaxation rate cancel in their ratio. Where relaxation dominates diffusion, this ratio fixes a finite local density as the common width tends to zero. Nonzero mixing splits the physical stop poles and removes the VIA pinch associated with equal unmixed energies. Taking the common width to zero at fixed splitting and then expanding for $|\omega_1-\omega_2|\ll T$ gives the same susceptibility, the coefficient of $\partial_t\alpha$ in the density. This zero-width limit of the local kernels is formal because at fixed $L_w$ it eventually violates $L_w^{-1}\ll\gamma$. The analytic stop kernel assumes well-separated poles, $\gamma_1+\gamma_2\ll|\omega_1-\omega_2|$, for the dominant momenta at each relevant point in the wall. Where these poles overlap appreciably, the full width dependence must be retained at the same order in wall gradients. Figure~\ref{fig:source_accuracy} illustrates how overlapping poles modify the near-resonance structure of the leading thermal source.

For chiral fermions with the thermal dispersions used in Section~\ref{sec:numerical}, the particle and hole energies entering the VIA source coincide only at zero momentum. As the common width tends to zero, the phase space for their overlap shrinks and the integrated VIA source vanishes. With the chiral self-energy used in the resummed benchmark, the thermal flux remains finite. Its coefficient first grows with the Dirac mass and then falls as the Fermi occupation becomes Boltzmann suppressed. At the wall center the spatial divergence probes the mass derivative of this coefficient, accounting for the sign change of the source.

At first order in wall gradients, the induced density is proportional to $\partial_t\alpha$ and is therefore localized where the phase varies across the moving wall. The VIA diffusion equation yields a tail of length $D/v_w$ in the symmetric phase. Whether the second-order resummed source can be incorporated consistently into a diffusion equation based on local chemical potentials remains a question.

Determining the left-handed fermion density that biases weak sphalerons requires following the charge from the wall into the symmetric phase. In future work, we will study propagation between the wall and a distant observation point, together with charge transfer among species through three-field interactions such as the Yukawa interaction of Section~\ref{sec:transfer}. We will also consider general mass mixing for arbitrary numbers of complex scalars and Weyl fermions. A possible extension is to evaluate the resummed current with temperature and fluid velocity profiles derived from stress--energy conservation for the wall and plasma~\cite{LaurentClineWall}.

\section*{Acknowledgments}
The author thanks the Department of Physics at the University of Massachusetts Amherst for financial support through a graduate assistantship.

The author developed the theoretical framework and takes responsibility for the calculations and conclusions. OpenAI Codex assisted with implementing and running numerical checks of the current and source kernels, preparing figures, and revising the manuscript.

\appendix

\section{Gradient expansion on the closed time path}
\label{app:gradient_derivation}
We derive the local gradient expansion for a scalar mass matrix and then use a single insertion with finite wall momentum to determine its range of validity. Consider a scalar two-point function with contour indices $a,b$. For $x$ later than $y$ along the contour, the Heisenberg-picture trace can be written in the interaction picture as
\begin{equation}
  \label{eq:thermalaveragefunction}
  \begin{aligned}
  \boldsymbol{G}^{ab}(x,y) &= \langle \mathcal{T}^c \boldsymbol{\Phi}_H^{a}(x) (\boldsymbol{\Phi}^\dagger_H)^b (y)\rangle \\
  &={\operatorname{Tr} \left[\boldsymbol{\rho}(t_0)\boldsymbol{\Phi}_H^{a}(x) (\boldsymbol{\Phi}^\dagger_H)^b (y)\right]} \\
  &={\operatorname{Tr} \left[\boldsymbol{\rho}(t_0) \boldsymbol{U}_I(t_0,x)\boldsymbol{\Phi}_I^{a}(x)\boldsymbol{U}_I(x,y)(\boldsymbol{\Phi}^\dagger_I)^b (y)\boldsymbol{U}_I(y,t_0)\right]} \\
  \end{aligned}
\end{equation}
Here $t_0$ is the initial time and $\boldsymbol U_I$ is the interaction-picture evolution operator. For a thermal initial state, the Hamiltonian on the imaginary-time branch is held fixed at its initial value. Then
\begin{equation}
  \boldsymbol{\rho}(t_0) = \frac{e^{-\beta \boldsymbol{{H}}(t_0)}}{\operatorname{Tr}e^{-\beta \boldsymbol{{H}}(t_0)}} = \frac{e^{-\beta \boldsymbol{H}_0} \boldsymbol{U}_I(t_0-i\beta,t_0)}{\operatorname{Tr}\left[e^{-\beta \boldsymbol{H}_0} \boldsymbol{U}_I(t_0-i\beta,t_0)\right]}
\end{equation}
and the trace in Eq.~\eqref{eq:thermalaveragefunction} becomes
\begin{equation}
  \begin{aligned}
   \frac{\operatorname{Tr} \left[e^{-\beta \boldsymbol{H}_0} \boldsymbol{U}_I(t_0-i\beta,t_0) \boldsymbol{U}_I(t_0,x)\boldsymbol{\Phi}_I^{a}(x)\boldsymbol{U}_I(x,y)(\boldsymbol{\Phi}^\dagger_I)^b (y)\boldsymbol{U}_I(y,t_0)\right]}{\operatorname{Tr}\left[e^{-\beta \boldsymbol{H}_0} \boldsymbol{U}_I(t_0-i\beta,t_0)\right]} \\
  \end{aligned}
\end{equation}
Inserting energy eigenstates of the free Hamiltonian $\boldsymbol H_0$ gives
\begin{equation}
  \begin{aligned}
   &\frac{\sum_n e^{-\beta E_n}\langle n| { \left[ \boldsymbol{U}_I(t_0-i\beta,t_0) \boldsymbol{U}_I(t_0,x)\boldsymbol{\Phi}_I^{a}(x)\boldsymbol{U}_I(x,y)(\boldsymbol{\Phi}^\dagger_I)^b (y)\boldsymbol{U}_I(y,t_0)\right]}|n\rangle}{{\sum_n e^{-\beta E_n}\langle n| \boldsymbol{U}_I(t_0-i\beta,t_0)|n\rangle}}.
  \end{aligned}
\end{equation}
For arbitrary contour ordering, the normalized trace is
\begin{equation}
  \begin{aligned}
  \langle \mathcal{T}^c \boldsymbol{\Phi}_H^{a}(x) (\boldsymbol{\Phi}^\dagger_H)^b (y)\rangle  &= \frac{\langle  \mathcal{T}^c \boldsymbol{\Phi}_I^{a}(x) (\boldsymbol{\Phi}^\dagger_I)^b (y) \exp\left( - i \int_c d^4 \omega \boldsymbol{\mathcal{H}}_I\right)\rangle}{\langle  \mathcal{T}^c \exp\left( - i \int_c d^4 \omega \boldsymbol{\mathcal{H}}_I\right)\rangle} \\
  &= \left\langle \mathcal{T}^c \boldsymbol{\Phi}_I^{a}(x) (\boldsymbol{\Phi}^\dagger_I)^b (y) \sum_{n=0}^{\infty} \frac{1}{n!}\left[- i\int_c d^4 \omega   \boldsymbol{\mathcal{H}}_I (\omega)\right]^n \right\rangle_{\text{con.}}
  \end{aligned}
\end{equation}
where $\boldsymbol{\mathcal H}_I$ denotes the interaction Hamiltonian density. For the nonderivative interactions considered here, $\boldsymbol{\mathcal H}_I=-\boldsymbol{\mathcal L}_{\rm int}$. The denominator cancels disconnected vacuum diagrams, and the remaining contractions use the free thermal state.

Choosing $\boldsymbol H_0$ to contain the kinetic terms gives
\begin{equation}
  \boldsymbol{\mathcal{H}}_I = \boldsymbol{\Phi}^\dagger \boldsymbol{M}^2(\omega)\boldsymbol{\Phi} - (\boldsymbol{J}^\dagger \boldsymbol{\Phi} + \text{h.c.}).
\end{equation}
Expand the mass matrix about the observation point,
\begin{equation}
  \boldsymbol{M}^2(\omega) =\boldsymbol{M}^2(x) + \partial_\mu \boldsymbol{M}^2(x) (\omega-x)^\mu + \frac{1}{2!}\partial_\mu \partial_\nu \boldsymbol{M}^2(x) (\omega-x)^\mu(\omega-x)^\nu + \cdots
\end{equation}
The constant mass and the homogeneous bath self-energy can now be included to all orders. In the notation of Eq.~\eqref{eq:inversezerothG}, the free massless propagator $\tilde{\boldsymbol G}^{(0)}(k)$ gives the propagator in the background frozen at $x$ through the geometric series
\begin{equation}
  \begin{aligned}
    \tilde{\boldsymbol{G}}^{(0)}(x,y) &= \int \frac{d^4 k}{(2\pi)^4} \tilde{\boldsymbol{G}}^{(0)}(k) \left[ \boldsymbol{I}+ \sum_{n=1}^\infty \left(\left(-i\tilde{\boldsymbol{M}}^2(x) - i\tilde{\boldsymbol{\Pi}}^{(0)}(k) \right) \tilde{\boldsymbol{G}}^{(0)}(k)\right)^n \right] e^{-ik(x-y)} \\
    &=\int \frac{d^4 k}{(2\pi)^4}\tilde{\boldsymbol{G}}^{(0)}(k) \frac{1}{1-\left(-i\tilde{\boldsymbol{M}}^2(x) - i\tilde{\boldsymbol{\Pi}}^{(0)}(k) \right) \tilde{\boldsymbol{G}}^{(0)}(k)}e^{-ik(x-y)} \\ 
    &= \int \frac{d^4 k}{(2\pi)^4} \frac{1}{\tilde{\boldsymbol{G}}^{(0)}(k)^{-1}+i\tilde{\boldsymbol{M}}^2(x) + i\tilde{\boldsymbol{\Pi}}^{(0)}(k) }e^{-ik(x-y)}.
  \end{aligned}
\end{equation}
We denote the resulting momentum-space propagator by $\tilde{\boldsymbol G}^{(0)}(k,x)$ and use it on both sides of every gradient insertion. With
$D_a=\partial/\partial k_a$, one gradient insertion gives
\begin{equation}
  \label{eq:first_order_gradient}
  \begin{aligned}
    &\tilde{\boldsymbol G}^{(1)}(x,y)\\
    &= -i\int d^4\omega\,
       \tilde{\boldsymbol G}^{(0)}(x,\omega)(\omega-x)^a
       \partial_a\tilde{\boldsymbol M}^2(x)\tilde{\boldsymbol G}^{(0)}(\omega,y)\\
    &= -i\int d^4\omega \int\frac{d^4k}{(2\pi)^4}\frac{d^4p}{(2\pi)^4}
       \tilde{\boldsymbol G}^{(0)}(k,x)e^{-ik(x-\omega)}(\omega-x)^a
       \partial_a\tilde{\boldsymbol M}^2(x)\tilde{\boldsymbol G}^{(0)}(p,x)e^{-ip(\omega-y)}\\
    &= \int\frac{d^4k}{(2\pi)^4}\,
       \big(D_a\tilde{\boldsymbol G}^{(0)}(k,x)\big)
       \partial_a\tilde{\boldsymbol M}^2(x)\tilde{\boldsymbol G}^{(0)}(k,x)e^{-ik(x-y)}.
  \end{aligned}
\end{equation}
The last line follows by integration by parts.
At second order there are two contributions, one insertion of the second mass derivative and a pair of ordered first-derivative insertions. Write $u=\omega-x$ and $v=\omega'-x$. At coincidence the same Fourier transform gives
\begin{equation}
\label{eq:second_order_gradient}
\begin{aligned}
 \tilde{\boldsymbol G}^{(2)}(x,x)
 &= -\frac{i}{2}\int d^4u\,u^au^b\,
 \tilde{\boldsymbol G}^{(0)}(x,x+u)\,
 \partial_a\partial_b\tilde{\boldsymbol M}^2(x)\,
 \tilde{\boldsymbol G}^{(0)}(x+u,x)\\
 &\quad+(-i)^2\int d^4u\,d^4v\,u^av^b\,
 \tilde{\boldsymbol G}^{(0)}(x,x+u)\,
 \partial_a\tilde{\boldsymbol M}^2(x)\,
 \tilde{\boldsymbol G}^{(0)}(x+u,x+v)\\
 &\qquad\qquad\times
 \partial_b\tilde{\boldsymbol M}^2(x)\,
 \tilde{\boldsymbol G}^{(0)}(x+v,x)\\
 &=\int\frac{d^4k}{(2\pi)^4}\Big[
 \frac{i}{2}\big(D_aD_b\tilde{\boldsymbol G}^{(0)}(k,x)\big)\,
 \partial_a\partial_b\tilde{\boldsymbol M}^2(x)\,
 \tilde{\boldsymbol G}^{(0)}(k,x)\\
 &\qquad\qquad-
 \big(D_a\tilde{\boldsymbol G}^{(0)}(k,x)\big)\,
 \partial_a\tilde{\boldsymbol M}^2(x)\,
 \tilde{\boldsymbol G}^{(0)}(k,x)\,
 \partial_b\tilde{\boldsymbol M}^2(x)\,
 \big(D_b\tilde{\boldsymbol G}^{(0)}(k,x)\big)\Big].
\end{aligned}
\end{equation}
For the scalar current the point-split derivative supplies $k^\mu$ on
the leftmost propagator before the momentum derivatives act, as in
Eq.~\eqref{eq:fojLz}. For fermions, the same expansion uses the resummed spinor propagator and Dirac mass vertex in Eq.~\eqref{eq:second_order_gradient}, followed by the left-handed spin trace with $-\bar\sigma^\mu/2$. For mixed derivatives, the terms with indices $(a,b)$ and $(b,a)$ combine to cancel the Taylor factor of $1/2$.

The same one-insertion expression determines how accurately a local gradient describes the wall. Keep the wall momentum $q$ before taking derivatives, and write the mass vertex as a Fourier transform centered at $x$,
\begin{equation}
  \label{eq:gradient_mass_insertion_fourier}
  \tilde{\boldsymbol M}^2(\omega)
  =\int\frac{d^4q}{(2\pi)^4}\,
  e^{-iq(\omega-x)}\tilde{\boldsymbol M}^2(q;x).
\end{equation}
The $\omega$ integral sets $k=p+q$, so a single mass vertex carries the kernel
\begin{equation}
  \label{eq:gradient_one_insertion_finite_q}
  \begin{aligned}
    \big[\tilde{\boldsymbol G}(x,y)\big]_{\text{one insertion}}
    &= -i\int\frac{d^4p}{(2\pi)^4}\frac{d^4q}{(2\pi)^4}\,e^{-ip(x-y)}\\
    &\qquad\times\tilde{\boldsymbol G}^{(0)}(p+q,x)
    \tilde{\boldsymbol M}^2(q;x)\tilde{\boldsymbol G}^{(0)}(p,x).
  \end{aligned}
\end{equation}
At fixed $p$ and $x$, the factor to expand is the shifted propagator $\tilde{\boldsymbol G}^{(0)}(p+q,x)$. Under the $q$ integral each factor $-iq_a$ differentiates the wall profile in Eq.~\eqref{eq:gradient_mass_insertion_fourier} with respect to $\omega^a$ at $\omega=x$. The constant mass is already included in the frozen propagator. The term linear in $q$ reproduces Eq.~\eqref{eq:first_order_gradient}, and successive powers give higher wall derivatives. For local observables we take $y\to x$ after any point-split derivatives.

At energies $|p^0|\sim T$, the thermal occupation factors vary on the scale $T$. A smaller scale can enter through the energy integral when a retarded and an advanced pole approach the contour from opposite sides. In the statistical component, $\boldsymbol G^>+\boldsymbol G^<=h(\boldsymbol G^r-\boldsymbol G^a)$ at zero bath chemical potentials, where $h(p^0)=\coth(p^0/2T)$ for scalars and $\tanh(p^0/2T)$ for fermions. Applying this identity to the propagator on either side of the insertion gives the retarded--advanced term
\[
 [h(p^0+q^0)-h(p^0)]\,
 \boldsymbol G^{(0),r}(p+q,x)\,\boldsymbol M^2(q;x)\,
 \boldsymbol G^{(0),a}(p,x),
\]
where $\boldsymbol M^2$ is the flavor block of the mass vertex. The two weights have opposite signs because the advanced part of the right statistical propagator enters with a minus sign.

Let $E_{i,\boldsymbol p}$ and $E_{j,\boldsymbol p}$ be the signed
energies of the retarded and advanced branches, with widths $\gamma_i$
and $\gamma_j$ held fixed over the shift. Write
$\Delta E_{ij}=E_{i,\boldsymbol p}-E_{j,\boldsymbol p}$ and
$\delta_q=q^0-(E_{i,\boldsymbol p+\boldsymbol q}-E_{i,\boldsymbol p})$.
At fixed $\boldsymbol p$ and $q$, collect the factors regular at the advanced pole in $\mathcal N(p^0)$. Taking the advanced residue in the $p^0$ integral of Eq.~\eqref{eq:gradient_one_insertion_finite_q} gives
\begin{equation}
 \label{eq:gradient_ra_pair}
 \begin{aligned}
 \left[\int\frac{dp^0}{2\pi}\,
 \frac{\mathcal N(p^0)}
 {(p^0+q^0-E_{i,\boldsymbol p+\boldsymbol q}+i\gamma_i)
  (p^0-E_{j,\boldsymbol p}-i\gamma_j)}\right]_{\mathrm{adv}}
 &=\mathcal N(E_{j,\boldsymbol p}+i\gamma_j)\,K_{ij}(q),\\
 K_{ij}(q)&=\frac{1}{\gamma_i+\gamma_j+i\Delta E_{ij}-i\delta_q}.
 \end{aligned}
\end{equation}
All wall-momentum dependence of the pole denominator enters through $\delta_q$, which vanishes at $q=0$. The fractional change from using its local value is exactly
\[
 \frac{|K_{ij}(q)-K_{ij}(0)|}{|K_{ij}(q)|}
 =\frac{|\delta_q|}{\sqrt{(\Delta E_{ij})^2+(\gamma_i+\gamma_j)^2}}.
\]
This ratio controls the Taylor expansion of the pole factor. The statistical weight difference already starts at first order in $q$, so the first correction to $K_{ij}(0)$ can enter a second-order observable.

For a smooth, steadily translating planar wall of thickness $L_w$, the characteristic $|\boldsymbol q|$ is of order $L_w^{-1}$, up to a factor fixed by the profile. Since $|q^0|\leq|\boldsymbol q|$ and the relativistic dispersion satisfies $|E_{i,\boldsymbol p+\boldsymbol q}-E_{i,\boldsymbol p}|\leq|\boldsymbol q|$, the shift $|\delta_q|$ is at most $2|\boldsymbol q|$. For poles separated by $|\Delta E_{ij}|\sim T$, the pole factor and the smooth thermal weights thus admit a local expansion for $L_w^{-1}\ll T$. For distinct but nearby branches, the expansion parameter retains both the real energy gap and the sum of the widths. For a same-branch pair, $\Delta E_{ii}=0$ and the poles are separated by $2\gamma_i$ in imaginary energy. If this term survives in the observable, $L_w^{-1}\ll\gamma_i$ is sufficient to expand it locally~\cite{PostmaRange}. Combining this locality condition with the narrow-width assumption gives the power counting
\[
 L_w^{-1}\ll\gamma_i\ll T\,.
\]

\section{Scalar propagator poles and statistical coefficients}
\label{app:boson_ctp_coefficients}
The mass-basis CTP propagator in
Eqs.~\eqref{eq:boson_compact_ctp} and \eqref{eq:boson_compact_spectral}
is assembled from the retarded, advanced, and lesser blocks
\begin{equation*}
  \begin{gathered}
    \widetilde{\boldsymbol G}^{(0)}_m \simeq
    \begin{pmatrix}
      \boldsymbol R+\boldsymbol L & -\boldsymbol L\\
      \boldsymbol R-\boldsymbol A+\boldsymbol L & \boldsymbol A-\boldsymbol L
    \end{pmatrix},\\[4pt]
    \boldsymbol R=\frac{i\boldsymbol{\mathcal N}_{+}}{\mathcal D_{+}},
    \qquad
    \boldsymbol A=\frac{i\boldsymbol{\mathcal N}_{-}}{\mathcal D_{-}},
    \qquad
    \boldsymbol L=\boldsymbol R\boldsymbol n-\boldsymbol n\boldsymbol A
                  -\boldsymbol R\boldsymbol{\mathcal K}\boldsymbol A.
  \end{gathered}
\end{equation*}
Each block is a $2\times2$ matrix in the mass indices.
We distinguish the parameters of the diagonal numerator zeros, labeled
$\mathrm N$, from those of the mixed propagator poles, labeled $\mathrm P$.
As in Section~\ref{sec:scalars}, we use the mixing matrix $\boldsymbol U$ and write $c=\cos\theta$, $s=\sin\theta$ and $\omega_i=(\boldsymbol k^2+m_i^2)^{1/2}$. Flavor sums run over $f=L,R$, with
$|U_{L1}|^2=|U_{R2}|^2=c^2$ and
$|U_{R1}|^2=|U_{L2}|^2=s^2$. The self-energy inputs
$\Pi_{h,f}$ and $\Gamma_f$ include their $O(g^2)$ scaling.

For $\sigma=+1$ (retarded) and $\sigma=-1$ (advanced), define
\begin{equation}
\label{eq:boson_compact_numerator_factor}
 F_{i\sigma}(k^0)=\prod_{\eta=\pm1}
 \left[k^0-\eta\Omega_i^{\mathrm N}
 +i\sigma\Gamma_i^{\mathrm N}
       \left(1-\frac{\eta\mu_i^{\mathrm N}}{\omega_i}\right)\right],
\end{equation}
\begin{equation}
\label{eq:boson_compact_pole_factor}
 P_{i\sigma}(k^0)=\prod_{\eta=\pm1}
 \left[k^0-\varepsilon_i-\eta\Omega_i^{\mathrm P}
 +i\sigma\Gamma_i^{\mathrm P}
       \left(1-\frac{\eta\mu_i^{\mathrm P}}{\omega_i}\right)\right].
\end{equation}
The spectral matrices are then
\begin{equation}
\label{eq:boson_compact_coefficient_matrices}
 \boldsymbol{\mathcal N}_\sigma=
 \begin{pmatrix}
  F_{2\sigma}&-e^{-i\alpha}\Pi^{\mathrm{mix}}_\sigma\\
  -e^{i\alpha}\Pi^{\mathrm{mix}}_\sigma&F_{1\sigma}
 \end{pmatrix},
 \qquad
 \mathcal D_\sigma=P_{1\sigma}P_{2\sigma}.
\end{equation}
For each pole pair, $\Omega_i^{\mathrm P}$ is half the separation of the
real pole energies, $\varepsilon_i$ is their real midpoint, and
$\Gamma_i^{\mathrm P}$ is the mean damping width. The parameter
$\mu_i^{\mathrm P}$ measures the difference between the positive- and
negative-frequency damping widths through
$\Gamma_i^{\mathrm P}(1\mp\mu_i^{\mathrm P}/\omega_i)$.
The numerator parameters have the analogous meaning for the diagonally
dressed channels before residual mixing. In particular, their mean real
midpoint is zero at the retained order.

The parameters $\mu_i^{\mathrm N}$ and $\mu_i^{\mathrm P}$ characterize
the difference between the damping widths at positive and negative frequencies. The statistical occupations remain
$n_f=n_B(k^0,\mu_f)$ with the physical flavor chemical potentials; the compact
spectral parameters do not replace $\mu_L$ and $\mu_R$ in the Bose functions.
The bare $\omega_i$ in the damping factors fixes the parameterization used
throughout the root expansion.

The numerator width is the flavor-projected diagonal width,
\begin{equation}
\label{eq:boson_numerator_width}
 \Gamma_i^{\mathrm N}=\sum_f |U_{fi}|^2\Gamma_f.
\end{equation}
The expressions for $\Omega_i^{\mathrm N}$ and $\mu_i^{\mathrm N}$ at the order required for the $O(g^4)$ roots are
\begin{equation}
\label{eq:boson_numerator_energy}
\begin{aligned}
 \Omega_i^{\mathrm N}={}&\omega_i
 +\frac{\sum_f|U_{fi}|^2\Pi_{h,f}}{2\omega_i}
 -\frac{\bigl(\sum_f|U_{fi}|^2\Pi_{h,f}\bigr)^2}{8\omega_i^3}\\
 &-\frac{(\Gamma_i^{\mathrm N})^2}{2\omega_i}
 +\frac{\bigl(\sum_f|U_{fi}|^2\Gamma_f\mu_f\bigr)^2}{2\omega_i^3},
\end{aligned}
\end{equation}
\begin{equation}
\label{eq:boson_numerator_chemical}
 \mu_i^{\mathrm N}=
 \frac{\sum_f|U_{fi}|^2\Gamma_f\mu_f}{\Gamma_i^{\mathrm N}}
 \left(1-\frac{\sum_f|U_{fi}|^2\Pi_{h,f}}{2\omega_i^2}\right).
\end{equation}
To leading order, $\mu_i^{\mathrm N}$ is a weighted average of the flavor chemical potentials, with weights proportional to $|U_{fi}|^2\Gamma_f$,
while the bracket accounts for the thermal energy correction. The numerator
energy contains the diagonal thermal mass, width, and chemical-potential
corrections without the additional mixing shift.

To express the mixing shifts in the denominator poles, define the following differences between flavors
\begin{equation}
\label{eq:boson_flavor_contrasts}
\begin{aligned}
 \Delta\Pi_h&=\Pi_{h,L}-\Pi_{h,R},\qquad
 \Delta\Gamma=\Gamma_L-\Gamma_R,\\
 \Delta(\Gamma\mu)&=\Gamma_L\mu_L-\Gamma_R\mu_R.
\end{aligned}
\end{equation}
With $j=3-i$, the pole parameters are
\begin{equation}
\label{eq:boson_pole_energy}
 \Omega_i^{\mathrm P}=\Omega_i^{\mathrm N}
 +\frac{c^2s^2\left[(\Delta\Pi_h)^2
      -4\omega_i^2(\Delta\Gamma)^2
      -4\{\Delta(\Gamma\mu)\}^2\right]}
       {2\omega_i(\omega_i^2-\omega_j^2)},
\end{equation}
\begin{equation}
\label{eq:boson_pole_width}
 \Gamma_i^{\mathrm P}=\Gamma_i^{\mathrm N}
 +\frac{2c^2s^2\Delta\Pi_h\Delta\Gamma}
       {\omega_i^2-\omega_j^2},
\end{equation}
\begin{equation}
\label{eq:boson_pole_midpoint}
 \varepsilon_i=
 \frac{4c^2s^2\Delta\Gamma\Delta(\Gamma\mu)}
      {\omega_i^2-\omega_j^2},
\end{equation}
\begin{equation}
\label{eq:boson_pole_chemical}
 \mu_i^{\mathrm P}=
 \frac{\displaystyle\Gamma_i^{\mathrm N}\mu_i^{\mathrm N}
 +\frac{2c^2s^2\Delta\Pi_h\Delta(\Gamma\mu)}
       {\omega_i^2-\omega_j^2}}
 {\displaystyle\Gamma_i^{\mathrm P}}.
\end{equation}
These are the additional effects of mixing on the diagonal thermal channels.
The real midpoint $\varepsilon_i$ is needed at nonzero chemical potentials:
the two real pole energies are generally
$\varepsilon_i\pm\Omega_i^{\mathrm P}$, rather than an exactly opposite pair.
The product $\Gamma_i^{\mathrm P}\mu_i^{\mathrm P}$ is fixed by the
numerator of Eq.~\eqref{eq:boson_pole_chemical}; its apparent width denominator
cancels in Eq.~\eqref{eq:boson_compact_pole_factor}. This parameterization
therefore reproduces the retained root polynomial and does not add a new
resummation. The displayed mixing corrections assume a fixed nonzero mass
splitting.

The same differences enter the mixing numerator and the statistical kernel.
The flavor absorptive factors and residual mixing self-energy are
\begin{equation}
\label{eq:boson_absorptive_factors}
 \mathcal A_f(k^0)=(k^0-\mu_f)\Gamma_f,\qquad f=L,R,
\end{equation}
\begin{equation}
\label{eq:boson_mixing_numerator}
 \Pi^{\mathrm{mix}}_\sigma(k^0)=cs
 \left[\Delta\Pi_h-2i\sigma(\mathcal A_L-\mathcal A_R)\right].
\end{equation}
In the same notation, the statistical kernel is
\begin{equation}
\label{eq:boson_statistical_kernel_compact}
 \boldsymbol{\mathcal K}=(n_R-n_L)
 \begin{pmatrix}
 4s^2\mathcal A_R&cs\,e^{-i\alpha}
       [2(\mathcal A_L+\mathcal A_R)+i\Delta\Pi_h]\\
 cs\,e^{i\alpha}[2(\mathcal A_L+\mathcal A_R)-i\Delta\Pi_h]
       &-4s^2\mathcal A_L
 \end{pmatrix}.
\end{equation}
The statistical correction is proportional to the difference $n_R-n_L$ between the Bose distributions. The off-diagonal self-energy depends on the differences between the two flavor self-energies. Equal occupations therefore eliminate
$\boldsymbol{\mathcal K}$ even when the off-diagonal self-energy is nonzero.

\section{Fermion propagator poles and statistical coefficients}
\label{app:fermion_ctp_coefficients}
The fermion CTP propagator in
Eqs.~\eqref{eq:fermion_compact_ctp} and \eqref{eq:fermion_compact_blocks}
is assembled from retarded, advanced, and lesser Weyl blocks at each
helicity $h=\pm1$,
\begin{equation*}
  \begin{gathered}
    \tilde{\boldsymbol S}^{(0)} \simeq
    \sum_{h=\pm1}
    \begin{pmatrix}
      \boldsymbol R_h+\boldsymbol L_h & -\boldsymbol L_h\\
      \boldsymbol R_h-\boldsymbol A_h+\boldsymbol L_h &
      \boldsymbol A_h-\boldsymbol L_h
    \end{pmatrix}
    \otimes\boldsymbol P^h,\\[4pt]
    \boldsymbol R_h=\frac{i\boldsymbol{\mathcal N}_{h,+}}{\mathcal D_{h,+}},
    \qquad
    \boldsymbol A_h=\frac{i\boldsymbol{\mathcal N}_{h,-}}{\mathcal D_{h,-}},
    \qquad
    \boldsymbol L_h=-\frac{n_{F,L}+n_{F,R}}{2}(\boldsymbol R_h-\boldsymbol A_h)
                   -\boldsymbol R_h\boldsymbol{\mathcal K}_F\boldsymbol A_h.
  \end{gathered}
\end{equation*}
For $\sigma=+1$ (retarded) and $\sigma=-1$ (advanced), the self-energy
coefficients give
\begin{equation}
  Z_I=1-a_I,\qquad
  c_I^\sigma=\frac{b_I-2i\sigma\Gamma_I}{Z_I},
  \qquad I=L,R.
\end{equation}
The numerator in Eq.~\eqref{eq:fermion_compact_blocks} is
\begin{equation}
  \begin{aligned}
    X_{L,h}^\sigma &= Z_L(k^0+hk-c_L^\sigma), \qquad
    X_{R,h}^\sigma = Z_R(k^0-hk-c_R^\sigma),\\
    \boldsymbol{\mathcal N}_{h,\sigma}
      &= \begin{pmatrix}
           m_D e^{-i\alpha} & X_{R,h}^\sigma\\
           X_{L,h}^\sigma & m_D e^{i\alpha}
         \end{pmatrix}.
  \end{aligned}
\end{equation}
For the denominator, define
\begin{equation}
  E=\sqrt{k^2+\frac{m_D^2}{Z_L Z_R}},\qquad
  \delta^\sigma=\frac{c_L^\sigma-c_R^\sigma}{2}.
\end{equation}
The factored denominator with pole shifts through $\mathcal O(g^4)$ is
\begin{equation}
  \label{eq:fermion_compact_poles}
  z_{h,\eta}^{\sigma}
    =\frac{c_L^\sigma+c_R^\sigma}{2}
      +\eta\left[E-\frac{hk\,\delta^\sigma}{E}
        +\frac{m_D^2(\delta^\sigma)^2}{2Z_L Z_R E^3}\right],
  \qquad
  \mathcal D_{h,\sigma}=Z_L Z_R\prod_{\eta=\pm1}(k^0-z_{h,\eta}^\sigma).
\end{equation}
Here $\eta=\pm1$, with $Z_I>0$ and $|\delta^\sigma|\ll E$.
The statistical kernel is
\begin{equation}
  \label{eq:fermion_statistical_kernel}
  \boldsymbol{\mathcal K}_F
    =2(n_{F,L}-n_{F,R})
     \begin{pmatrix} 0 & \Gamma_R\\ -\Gamma_L & 0 \end{pmatrix}.
\end{equation}

At finite $m_D$, the exact denominator $X_{L,h}^\sigma X_{R,h}^\sigma-m_D^2$
is quadratic in $k^0$ for each helicity. Its two roots tend to
$k^0=\eta\sqrt{k^2+m_D^2}$ as the self-energies are turned off, with
$\eta=+1$ for the particle branch and $\eta=-1$ for the antiparticle
branch. Using the retarded poles of Eq.~\eqref{eq:fermion_compact_poles},
we define the excitation energies and damping rates by
\begin{equation}
  \varepsilon_{h,\eta}(k)=\eta\,\operatorname{Re}z^+_{h,\eta}(k),
  \qquad
  \gamma_{h,\eta}(k)=-\operatorname{Im}z^+_{h,\eta}(k).
\end{equation}
For real self-energy parameters the advanced poles satisfy
$z^-_{h,\eta}=(z^+_{h,\eta})^*$.
As $k\to0$, the poles become independent of helicity and the rest energies are
\begin{equation}
  m_\eta^{\mathrm{pole}}\equiv\lim_{k\to0}\varepsilon_{h,\eta}(k)
    = E(0)
      +\frac{\eta}{2}\operatorname{Re}(c_L^++c_R^+)
      +\frac{\operatorname{Re}\!\left[(\delta^+)^2\right]}{2E(0)}.
\end{equation}
This expression uses Eq.~\eqref{eq:fermion_compact_poles} in the regime
$|\delta^+|\ll E(0)=m_D/\sqrt{Z_LZ_R}$, with the pole shifts small enough
to preserve the signs of the two branches.

The Dirac mass allows both poles to contribute to each
chiral propagator. In the limit $m_D\to0$ at fixed $k>0$, each chiral
propagator reduces to one linear denominator per helicity. In the sign
convention of Appendix~B of
Ref.~\cite{PostmaCancel}, these denominators are
\begin{equation}
  D_{I,\pm}(k^0,k)=Z_I(k^0\mp k)-b_I+2i\Gamma_I.
\end{equation}
The unique roots of $D_{I,+}$ and $D_{I,-}$
connect to $+k$ and $-k$, respectively, and both have damping rate
$2\Gamma_I/Z_I$. These are the particle and antiparticle poles of the chiral propagator.

Additional poles can arise from the energy dependence of the self-energy.
For example, in the chiral limit with widths neglected, the
hard-thermal-loop (HTL) denominators of Eq.~(B.13) in Ref.~\cite{PostmaCancel} approach
\begin{equation}
  D_{I,\pm}^{\mathrm{HTL}}(k^0,0)=k^0-\frac{m_{T,I}^2}{k^0},
\end{equation}
where $m_{T,I}$ is the thermal fermion mass. Each denominator now has
roots at $k^0=\pm m_{T,I}$. At nonzero $k$, the positive-energy roots of
$D_{I,+}^{\mathrm{HTL}}$ and $D_{I,-}^{\mathrm{HTL}}$ give the normal
particle and collective hole (plasmino) branches, respectively. The
negative-energy roots describe the antihole and antiparticle, respectively.

At fixed background and fixed $a_I$, $b_I$ and $\Gamma_I$, the chemical
potentials enter only through the Fermi occupations in the Wightman
functions, including the difference in
Eq.~\eqref{eq:fermion_statistical_kernel}. The retarded and advanced
propagators and their poles are thus $\mu_I$ independent before the
$\mathcal O(g^4)$ expansion. The boson ansatz uses $k^0-\mu_I$ in the
absorptive self-energy to cancel the Bose pole at $k^0=\mu_I$. The Fermi
occupation is finite there, so we use a constant fermion width.
In a microscopic calculation, the medium occupations can give the
self-energy, and hence $a_I$, $b_I$ and $\Gamma_I$, a chemical-potential
dependence that shifts the dispersions and damping rates. See
Ref.~\cite{Morales} for a
finite-density calculation of fermion dispersions.

\section{Second-order sources for scalar mixing}
\label{app:boson_second_order_sources}

After integration over $k^0$,
Eqs.~\eqref{eq:masssource2nd} and \eqref{eq:decaysource2nd} give the
following kernels at $\mu_L=\mu_R=0$ through $\mathcal O(g^0)$.
For $\sigma=0,3$, write $\partial_\sigma=\partial/\partial x^\sigma$.
The subscripts 2 and 3 count the propagator factors in each kernel. In the decay source the mixed correlator $D$ counts as one factor.
\begin{subequations}
\label{eq:boson_explicit_second_order_sources}
\begin{align}
 \left[{}^{\text{CPV}(2)}S_L^m(x)\right]_\sigma
   &=\int\frac{d^3k}{(2\pi)^3}\left[
     \partial_\sigma(v^4\partial_\sigma\alpha)\,\mathcal M_{\sigma,2}
     +2v^2(\partial_\sigma v^2)(\partial_\sigma\alpha)\,
       \mathcal M_{\sigma,3}\right],\\
 \left[{}^{\text{CPV}(2)}S_L^d(x)\right]_\sigma
   &=\int\frac{d^3k}{(2\pi)^3}\Big[
     \big(v^2\partial_\sigma^2\alpha
        +2(\partial_\sigma v^2)(\partial_\sigma\alpha)\big)
         \mathcal D_{\sigma,2}\notag\\
   &\hspace{8em}
     +2(\partial_\sigma v^2)(\partial_\sigma\alpha)
       \mathcal D_{\sigma,3}\Big].
\end{align}
\end{subequations}
Each source is the sum of its temporal and spatial contributions.
We use the widths $\Gamma_i^{\mathrm N}$ and the width difference $\Delta\Gamma$
from Eqs.~\eqref{eq:boson_numerator_width} and \eqref{eq:boson_flavor_contrasts}.
In the pole sums $i=1,2$ and $j=3-i$. Write
\begin{equation}
 \Delta=\omega_1^2-\omega_2^2,\qquad
 H_i=\coth\!\left(\frac{\omega_i}{2T}\right).
\end{equation}
Primes on $H_i$ denote derivatives with respect to $\omega_i$ at fixed $T$.

\subsection{Mass-mixing source}

The two-propagator kernels are
\begin{subequations}
\label{eq:boson_mass_source_2p}
\begin{align}
 \mathcal M_{0,2}
   &=-\sum_{i=1}^2
       \left[
        \frac{(-1)^{i-1}(3\omega_i^2+\omega_j^2)H_i}
             {\omega_i\Delta^3}
        +\frac{s_\theta^2c_\theta^2(\Delta\Gamma)^2 H_i'}
              {(\Gamma_i^{\mathrm N})^2\Delta^2}
       \right],\\
 \mathcal M_{3,2}
   &=-\frac{1}{\Delta^2}\sum_{i=1}^2\frac{H_i}{\omega_i}
      +\frac{8T}{\Delta^3}
       \log\!\left(\frac{\sinh(\omega_1/2T)}{\sinh(\omega_2/2T)}\right).
\end{align}
\end{subequations}
The three-propagator kernels are
\begin{subequations}
\label{eq:boson_mass_source_3p}
\begin{align}
 \mathcal M_{0,3}
   &=s_\theta c_\theta v^2\sum_{i=1}^2\Bigg[
      \frac{(-1)^{i-1}(35\omega_i^4+14\omega_i^2\omega_j^2-\omega_j^4)H_i}
           {2\omega_i^3\Delta^4}
      -\frac{(-1)^{i-1}s_\theta^2c_\theta^2(\Delta\Gamma)^2 H_i''}
           {2(\Gamma_i^{\mathrm N})^2\omega_i\Delta^2}
      \notag\\
   &\quad+\frac{\Delta\Gamma H_i'}
                    {2\Gamma_i^{\mathrm N}\omega_i^2\Delta^3}
       \Bigg((1-2|U_{Li}|^2)
              \left[5\omega_i^2+\omega_j^2
                 +\frac{2s_\theta^2c_\theta^2(\Delta\Gamma)^2\omega_i^2}
                        {(\Gamma_i^{\mathrm N})^2}\right]
          +\frac{s_\theta^2c_\theta^2\Delta\Gamma}{\Gamma_i^{\mathrm N}}
                  (5\omega_i^2-\omega_j^2)\Bigg)
     \Bigg],\\
 \mathcal M_{3,3}
   &=s_\theta c_\theta v^2
       \Bigg[
       -\frac{48T}{\Delta^4}
        \log\!\left(\frac{\sinh(\omega_1/2T)}{\sinh(\omega_2/2T)}\right)
       \notag\\
   &\qquad
       +\sum_{i=1}^2\left[
        \frac{6H_i}{\omega_i\Delta^3}
        -\frac{(-1)^{i-1}}{2\Delta^2}
          \left(\frac{H_i'}{\omega_i^2}-\frac{H_i}{\omega_i^3}\right)
       \right]\Bigg].
\end{align}
\end{subequations}

\subsection{Decay source}

The two-propagator kernels are
\begin{subequations}
\label{eq:boson_decay_source_2p}
\begin{align}
 \mathcal D_{0,2}
    &=\frac{s_\theta c_\theta\Gamma_L\Delta\Gamma}{\Delta}
      \sum_{i=1}^2\frac{|U_{Li}|^2 H_i'}{(\Gamma_i^{\mathrm N})^2},\\
 \mathcal D_{3,2}
    &=0.
\end{align}
\end{subequations}
The three-propagator decay kernel also vanishes for spatial gradients.
\begin{subequations}
\label{eq:boson_decay_source_3p}
\begin{align}
 \mathcal D_{0,3}
    &=s_\theta^2c_\theta^2 v^2\Gamma_L\sum_{i=1}^2
       \Bigg[
        \frac{(-1)^{i-1}|U_{Li}|^2\Delta\Gamma H_i''}
             {2(\Gamma_i^{\mathrm N})^2\omega_i\Delta}
        \notag\\
    &\quad -\frac{H_i'}{2\Gamma_i^{\mathrm N}\omega_i^2\Delta^2}
        \left(3\omega_i^2+\omega_j^2
          +\frac{|U_{Li}|^2\Delta\Gamma}{\Gamma_i^{\mathrm N}}
               (5\omega_i^2-\omega_j^2)
          +\frac{2|U_{Li}|^2(1-2|U_{Li}|^2)(\Delta\Gamma)^2\omega_i^2}
                 {(\Gamma_i^{\mathrm N})^2}\right)
       \Bigg],\\
 \mathcal D_{3,3}&=0.
\end{align}
\end{subequations}

\section{Second-order sources for a Dirac fermion}
\label{app:fermion_second_order_sources}

After integration over $k^0$,
Eqs.~\eqref{eq:fermionmasssource2nd} and \eqref{eq:fermiondecaysource2nd}
give the following kernels at $\mu_L=\mu_R=0$ through $\mathcal O(g^0)$.
\begin{subequations}
\begin{align}
 \left[{}^{\mathrm{CPV}(2)}S_L^m\right]_\sigma
   &=\int\frac{d^3 k}{(2\pi)^3}\Big[
      \partial_\sigma(m_D^2\partial_\sigma\alpha)\,\mathcal M_{\sigma,2}^F
       \notag\\
   &\hspace{5em}
      +2m_D^2(\partial_\sigma m_D)(\partial_\sigma\alpha)
        \mathcal M_{\sigma,3}^F\Big],\\
 \left[{}^{\mathrm{CPV}(2)}S_L^d\right]_\sigma
   &=\int\frac{d^3 k}{(2\pi)^3}\Big[
      \big(m_D\partial_\sigma^2\alpha
         +2(\partial_\sigma m_D)(\partial_\sigma\alpha)\big)
         \mathcal D_{\sigma,2}^F\notag\\
   &\hspace{5em}
      +2m_D(\partial_\sigma m_D)(\partial_\sigma\alpha)
         \mathcal D_{\sigma,3}^F\Big].
\end{align}
\end{subequations}
The indices $\sigma=0$ and $\sigma=3$ select the temporal and spatial contributions, respectively. The subscripts 2 and 3 count propagator factors as in Appendix~\ref{app:boson_second_order_sources}.
Write
\begin{equation}
 \begin{gathered}
 H=\tanh\!\left(\frac{\omega}{2T}\right),\qquad
 H'=\frac{dH}{d\omega},\qquad
 \Gamma_\pm=\Gamma_L\pm\Gamma_R,\qquad
 Q=4\Gamma_L\Gamma_R k^2+\Gamma_+^2m_D^2,\\
 \mathcal D_{m_D}=\partial_{m_D}+\frac{m_D}{\omega}\partial_\omega .
 \end{gathered}
\end{equation}
The derivative $\mathcal D_{m_D}$ includes the mass dependence of $\omega$
at fixed $k$, $T$, and $\Gamma_{L,R}$.

The temporal kernels are
\begin{subequations}
\begin{align}
 \mathcal M_{0,2}^F
   &=\frac{H}{2\omega^3}
    +\frac{m_D^2\Gamma_-^2 H'}{2\omega^2 Q^2}
       \left[2(\Gamma_L^2+\Gamma_R^2)\omega^2
             -m_D^2\Gamma_-^2\right],\\
 \mathcal D_{0,2}^F
   &=-\frac{m_D\Gamma_L\Gamma_- H'}{Q^2}
           \left(\Gamma_+^2m_D^2+4\Gamma_R^2k^2\right),\\
 \mathcal D_{0,3}^F
   &=-\frac{m_D H}{2T\omega}\mathcal D_{0,2}^F
     +\frac{m_D^2\Gamma_L\Gamma_+^2 H'}{Q^3}
       \Big[m_D^2\Gamma_-^2(\Gamma_L+4\Gamma_R)\notag\\
   &\hspace{12em}
             +4\Gamma_R^2(2\Gamma_L-\Gamma_R)\omega^2\Big],\\
 \mathcal M_{0,3}^F
   &=\frac12\mathcal D_{m_D}\mathcal M_{0,2}^F
      +\frac{m_D\mathcal D_{m_D}\mathcal D_{0,2}^F-\mathcal D_{0,2}^F}
             {2m_D^2}
      -\frac{\mathcal D_{0,3}^F}{m_D}.
\end{align}
\end{subequations}
The spatial kernels are
\begin{equation}
 \mathcal M_{3,2}^F=\frac{H'}{2\omega^2}-\frac{H}{2\omega^3},
 \qquad
 \mathcal M_{3,3}^F=\frac12\mathcal D_{m_D}\mathcal M_{3,2}^F,
 \qquad
 \mathcal D_{3,2}^F=\mathcal D_{3,3}^F=0.
\end{equation}

\section{Local source and mass-basis flavor transport}
\label{app:mass_basis_transport}
In the flavor basis the scalar mass-mixing source is a local term in the continuity equation. In the mass basis a varying mixing matrix can generate off-diagonal coherence already at first order in gradients. To relate these statements, we first express the flavor source and current in terms of mass-basis correlators. We then evaluate the mass source for a collisionless, phase-varying wall using the transport equations of Refs.~\cite{Flavor,Oscillations}. With $m^2_{LR}=v^2e^{-i\alpha}$ and $m^2_{RL}=(m^2_{LR})^*$, the mass commutator~\cite{PostmaCancel} gives
\begin{align}
 S_L^m(x)
 &= -\frac{i}{2}\int\frac{d^4k}{(2\pi)^4}
       [\boldsymbol M^2,\boldsymbol G^>+\boldsymbol G^<]_{LL}
       \label{eq:flsource1}\\
 &=\operatorname{Im}\int\frac{d^4k}{(2\pi)^4}
       m^2_{LR}(G^>_{RL}+G^<_{RL}) .
\end{align}
This is the mass term in Eq.~\eqref{eq:EOMsource}. For a two-point function with midpoint $x$ and separation $r$, the basis rotations act at $x+r/2$ and $x-r/2$. Their Wigner transform through first order in gradients is
\begin{align}
 \boldsymbol G^\lambda(k,x)
 &=\boldsymbol U\boldsymbol G_m^\lambda\boldsymbol U^\dagger
 -\frac{i}{2}\partial_{k_\mu}
 \big[(\partial_\mu\boldsymbol U)\boldsymbol G_m^\lambda\boldsymbol U^\dagger
 -\boldsymbol U\boldsymbol G_m^\lambda(\partial_\mu\boldsymbol U^\dagger)\big]
 +\mathcal O(\partial^2).
\end{align}
The mass source involves the momentum integral of this expression. At coincident points the rotations are evaluated at the same position, so the local commutator takes the form
\begin{equation}
 S_L^m(x)=-\frac{i}{2}\int\frac{d^4k}{(2\pi)^4}
  \big(\boldsymbol U[\boldsymbol m^2,\boldsymbol G_m^>+\boldsymbol G_m^<]
          \boldsymbol U^\dagger\big)_{LL}.
 \label{eq:flsource2}
\end{equation}
For the current, the derivatives must act on the two endpoints before the coincidence limit. The point-split definition of the left flavor current is
\begin{align}
 j_L^\mu(x)
 &= i\big\langle:\Phi_L^\dagger\partial^\mu\Phi_L
          -(\partial^\mu\Phi_L^\dagger)\Phi_L:\big\rangle \notag\\
 &=\frac{i}{2}\lim_{y\to x}(\partial_x^\mu-\partial_y^\mu)
       \sum_{\lambda=<,>}G^\lambda_{LL}(x,y).
\end{align}
Substituting $\boldsymbol G^\lambda(x,y)=\boldsymbol U(x)\boldsymbol G_m^\lambda(x,y)\boldsymbol U^\dagger(y)$ gives two contributions. Differentiation of $\boldsymbol G_m^\lambda$ yields the rotated mass-basis current. Differentiation of the endpoint rotations introduces the connection $\boldsymbol\Sigma^\mu=\boldsymbol U^\dagger\partial^\mu\boldsymbol U$. The resulting matrix identity is
\begin{equation}
 \boldsymbol j^\mu=
  \boldsymbol U\left[\boldsymbol j_m^\mu
    +\frac{i}{2}\sum_{\lambda=<,>}
       \{\boldsymbol\Sigma^\mu,\boldsymbol G_m^\lambda(x,x)\}\right]
   \boldsymbol U^\dagger .
 \label{eq:vrsource}
\end{equation}
Equation~(71) of Ref.~\cite{Flavor} projects $\boldsymbol U(\boldsymbol f_m-\bar{\boldsymbol f}_m)\boldsymbol U^\dagger$ onto the left flavor to define the density. The relation following Eq.~(42) of Ref.~\cite{Oscillations} similarly gives the rotated quasiparticle current $\boldsymbol U\boldsymbol j_m^\mu\boldsymbol U^\dagger$. When $\boldsymbol U$ varies, the flavor current generally requires the $\boldsymbol\Sigma^\mu$ term in Eq.~\eqref{eq:vrsource}. Taking the divergence of the full point-split current gives $\Box_x-\Box_y$, which the collisionless field equations reduce to the mass commutator in Eq.~\eqref{eq:flsource2}.

In the quasiparticle conventions of Ref.~\cite{Oscillations}, $\boldsymbol f_m$ and $\bar{\boldsymbol f}_m$ are the particle and antiparticle density matrices and $\bar\omega_{\boldsymbol k}=(\omega_1+\omega_2)/2$. The current then reads
\begin{equation}
 \boldsymbol j^\mu =
 \boldsymbol U\int\frac{d^3 k}{(2\pi)^3\bar\omega_{\boldsymbol k}}
 \left[k^\mu(\boldsymbol f_m-\bar{\boldsymbol f}_m)
   +\frac{i}{2}\{\boldsymbol\Sigma^\mu,
      \boldsymbol 1+\boldsymbol f_m+\bar{\boldsymbol f}_m\}\right]
 \boldsymbol U^\dagger .
\end{equation}
Only off-diagonal coherence contributes to the mass commutator. After relabeling the antiparticle momentum, the equal-time correlator carries the weight $(\boldsymbol 1+\boldsymbol f_m+\bar{\boldsymbol f}_m)/\bar\omega_{\boldsymbol k}$. Using $m_1^2-m_2^2=2\bar\omega_{\boldsymbol k}(\omega_1-\omega_2)$ in Eq.~\eqref{eq:flsource2} gives
\begin{equation}
\label{eq:vrsource2}
    \begin{aligned}
        S_L^m(x) &= \frac{-i}{2}\int\frac{d^4k}{(2\pi)^4}\left( \boldsymbol U[\boldsymbol m^2,\boldsymbol G_m^< +\boldsymbol G_m^>]\boldsymbol U^\dagger \right)_{LL}\\
        &= -i\int\frac{d^3k}{(2\pi)^3}\left(\boldsymbol U[\boldsymbol \omega_{\mathbf{k}},\boldsymbol f_m+\bar{\boldsymbol f}_m]\boldsymbol U^\dagger \right)_{LL}\\
        &= 2\cos\theta \sin\theta \int\frac{d^3k}{(2\pi)^3} (\omega_1 - \omega_2) \operatorname{Re}\left[i e^{i\alpha}(f_{12}+\bar f_{12})\right]
    \end{aligned}
\end{equation}

To determine which part of this coherence contributes to $S_L^m$, consider a stationary planar wall. In the collisionless limit, the scalar density-matrix equation of Cirigliano, Lee and Tulin [Eq.~(39a) of Ref.~\cite{Oscillations}] gives
\begin{align}
    v_{\rm rel}\partial_z f_{12} &= -i(\omega_1-\omega_2)f_{12} - v_{\rm rel}(\Sigma_{11}-\Sigma_{22})f_{12} + v_{\rm rel}\Sigma_{12}(f_{11}-f_{22})\label{eq:kinetic1}\\
    v_{\rm rel}\partial_z f_{11} &= -v_{\rm rel}(\Sigma_{12}f_{21} - \Sigma_{21}f_{12}) \label{eq:kinetic2}\\
    v_{\rm rel}\partial_z f_{22} &= -v_{\rm rel}(\Sigma_{21}f_{12} - \Sigma_{12}f_{21}) \label{eq:kinetic3}
\end{align}
Here $v_{\rm rel}$ is the longitudinal quasiparticle velocity relative to the wall at momentum $\boldsymbol k$, and $\boldsymbol\Sigma=\boldsymbol U^\dagger\partial_z\boldsymbol U$.

Equations~\eqref{eq:kinetic1}--\eqref{eq:kinetic3} retain the kinetic operator through first order in wall gradients. They suffice to test for a first-order mass source. At fixed mass splitting their iteration generates some second-order terms, such as $v_{\rm rel}\partial_z f_{12}^{(1)}$. A complete second-order coefficient also requires the next order of the Wigner expansion of the operator, including terms of the form $i\{\partial_{x^\mu}(k\cdot\boldsymbol\Sigma),\partial_{k_\mu}\boldsymbol G_m^{\gtrless}\}$, where the derivative of $\boldsymbol\Sigma$ contains two gradients.

Take a smooth wall in which only $\alpha(z)$ varies, so $\Sigma_{12}=i\cos\theta\sin\theta\,e^{-i\alpha}\partial_z\alpha$. For an incoming mode with $v_{\rm rel}>0$, take equilibrium populations and zero coherence far upstream. A small phase perturbation induces $f_{12}$ at linear order in its amplitude. In Eqs.~\eqref{eq:kinetic2} and~\eqref{eq:kinetic3}, the induced coherence multiplies the off-diagonal connection, so the change in population is quadratic in that amplitude. Keeping $f_{ii}=n_B(\omega_i)$ in Eq.~\eqref{eq:kinetic1} therefore gives the leading coherence,
\begin{align}
 f_{12}(z) &=
 \big[n_B(\omega_1)-n_B(\omega_2)\big]
 \int_{-\infty}^{z} dz'\,\Sigma_{12}(z')\notag\\
 &\quad\times
 \exp\!\left[2i\sin^2\theta\,\big(\alpha(z')-\alpha(z)\big)
       +i\frac{\omega_1-\omega_2}{v_{\rm rel}}(z'-z)\right].
 \label{eq:particle_streaming}
\end{align}
In the adiabatic limit $L_{\rm osc}=2\pi|v_{\rm rel}/(\omega_1-\omega_2)|\ll L_w$, integration by parts gives the leading local response from the upper endpoint of Eq.~\eqref{eq:particle_streaming},
\begin{equation}
 f_{12}^{(1)}(z)=-i\Sigma_{12}(z)\,
 v_{\rm rel}\frac{n_B(\omega_1)-n_B(\omega_2)}{\omega_1-\omega_2}.
 \label{eq:f12firstorder}
\end{equation}
The antiparticle equation reverses the oscillation phase. Its first-order endpoint term cancels Eq.~\eqref{eq:f12firstorder}, so $f_{12}^{(1)}+\bar f_{12}^{(1)}=0$ at each momentum in Eq.~\eqref{eq:vrsource2}.

Away from the adiabatic limit we retain the full oscillation phase in Eq.~\eqref{eq:particle_streaming}. To linear order in the phase perturbation, the background phase factor multiplying $\partial_{z'}\alpha(z')$ can be evaluated as $e^{-i\alpha(z)}$. The sum of the particle and antiparticle solutions is then
\begin{align}
 f_{12}(z)+\bar f_{12}(z)
 &=i e^{-i\alpha(z)}\sin(2\theta)
 \big[n_B(\omega_1)-n_B(\omega_2)\big]\notag\\
 &\quad\times\int_{-\infty}^{z} dz'\,\partial_{z'}\alpha(z')
       \cos\!\left[\frac{\omega_1-\omega_2}{v_{\mathrm{rel}}}(z'-z)\right].
 \label{eq:conventionalVR}
\end{align}
The oscillatory integral can be used to bound the mass source for any ratio $L_{\rm osc}/L_w$. Let $q=(\omega_1-\omega_2)/v_{\rm rel}$ and denote $z$ derivatives by primes. For $q\ne0$, two integrations by parts in Eq.~\eqref{eq:conventionalVR} give
\begin{align}
 \int_{-\infty}^{z} dz'\,\alpha'(z')\cos[q(z'-z)]
 &=-\frac{1}{q}\int_{-\infty}^{z} dz'\,\alpha''(z')\sin[q(z'-z)]\notag\\
 &=\frac{1}{q^2}\left\{\alpha''(z)
   -\int_{-\infty}^{z} dz'\,\alpha'''(z')\cos[q(z'-z)]\right\}.
 \label{eq:coherence_ibp}
\end{align}
The derivatives of the phase vanish far upstream. At the upper endpoint the sine vanishes on the first integration and the cosine yields $\alpha''(z)/q^2$ on the second.

In the mass source, the splitting from the commutator in Eq.~\eqref{eq:vrsource2} combines with the thermal occupation difference in Eq.~\eqref{eq:conventionalVR} to give
\begin{equation}
 \frac{(\omega_1-\omega_2)[n_B(\omega_1)-n_B(\omega_2)]}{q^2}
 = v_{\rm rel}^2\,\frac{n_B(\omega_1)-n_B(\omega_2)}{\omega_1-\omega_2}.
\end{equation}
As both energies approach $\omega>0$, the difference quotient tends to $n_B'(\omega)$ and this factor tends to $v_{\rm rel}^2n_B'(\omega)$. The expression in braces in Eq.~\eqref{eq:coherence_ibp} is bounded in magnitude by $|\alpha''(z)|+\int_{-\infty}^{z}dz'\,|\alpha'''(z')|$. For a smooth wall of fixed shape and phase excursion, this bound scales as $L_w^{-2}$ for every $qL_w$. Thus the projected mass source in this example vanishes at first order, even when the oscillation length is comparable to or exceeds the wall thickness. This agrees with the local flavor-basis result of Section~\ref{sec:massmixingsource}. The bound applies to the projected mass source at linear order in the phase excursion, with fixed masses and mixing angle. Transported flavor densities in the presence of collisions, including the resonant diffusion studied in Ref.~\cite{Oscillations}, require additional kinetic information.

\phantomsection
\addcontentsline{toc}{section}{\refname}
\providecommand{\href}[2]{#2}\begingroup\raggedright\endgroup
\end{document}